\documentclass[abstracton,12pt,english]{scrartcl}
\KOMAoptions{titlepage=false}
\usepackage[authoryear]{natbib} 
\usepackage{amsmath}
\usepackage{amssymb,latexsym}
\usepackage{graphicx}
\usepackage{calc}
\usepackage[usenames]{color}
\usepackage{subfig}
\usepackage{tabu}
\usepackage{setspace}
\usepackage{times}

\usepackage{geometry}
\definecolor{darkred}{rgb}{1.0,1.0,0.749}
\definecolor{red}{rgb}{1.0,0.0,0.0}
\definecolor{lightred}{rgb}{0.843,0.1,0.11}
\definecolor{white}{rgb}{0.843,0.1,0.11}
\definecolor{lightblue}{rgb}{0.843,0.1,0.11}
\definecolor{blue}{rgb}{0.843,0.1,0.11}
\definecolor{darkblue}{rgb}{0.843,0.1,0.11}
\definecolor{green}{rgb}{0.13725, 0.58039, 0.11373}

\newcommand{\erf}{\mathrm{erf}}
\newcommand{\pd}{\partial}

\newcommand{\bc}{\begin{center}}
\newcommand{\ec}{\end{center}}
\newcommand{\bd}{\begin{description}}
\newcommand{\ed}{\end{description}}
\newcommand{\ben}{\begin{enumerate}}
\newcommand{\een}{\end{enumerate}}
\newcommand{\bi}{\begin{itemize}}
\newcommand{\ei}{\end{itemize}}
\newcommand{\be}{\begin{equation}}
\newcommand{\ee}{\end{equation}}
\newcommand{\bal}{\begin{align}}
\newcommand{\eal}{\end{align}}

\usepackage[left,displaymath,mathlines]{lineno}

\def\mrm#1{\mathrm{#1}}
\def\fig#1{Fig.\,\ref{#1}}

\def\sec#1{Sec.\,\ref{#1}}
\def\eq#1{Eq.\,(\ref{#1})}

\def\fig#1{Fig.\,\ref{#1}}
\begin{document}


\title{The effect of spatial scales on the reproductive fitness of plant pathogens}






\author{Alexey Mikaberidze$^{*,1}$, \\
Christopher C. Mundt$^{2}$,\\
Sebastian Bonhoeffer$^{1}$}

\publishers{

\begin{normalsize}

\begin{flushleft}

$^{*}$alexey.mikaberidze@env.ethz.ch, Institute of Integrative Biology, ETH Zurich, CHN H 75.1, Universitaetstrasse 16,
8092, Zurich, phone: +41 44 632 26 02\\

$^1$Institute of Integrative Biology, ETH Zurich\\

$^2$Department of Botany and Plant Pathology, Oregon State University

 \vspace{0.5cm}


\vspace{0.3cm}






\end{flushleft}

\end{normalsize}
}

\date{}

\maketitle

\newpage


\begin{flushleft}

\begin{abstract}

    Plant diseases often cause serious yield losses in
    agriculture. A pathogen's reproductive fitness can be quantified
    by the basic reproductive number, $R_0$. Since pathogen
    transmission between host plants depends on the spatial separation
    between them, $R_0$ is strongly influenced by the spatial scales
    of pathogen dispersal and the spatial scales of the host
    population.

  We propose a novel method to estimate the basic reproductive number
  as a function of the size of a field planted with crops and its
  aspect ratio. This approach is based on measurements of disease
  gradients and uses a spatially explicit population dynamical
  model. 

  The basic reproductive number was found to increase with the
  field size at small field sizes and to saturate to a constant value
  at large field sizes. It reaches a maximum in square fields and
  decreases as the field becomes elongated. This pattern appears to be
  quite general: it holds for dispersal kernels that decrease
  exponentially or faster as well as for “fat-tailed” dispersal
  kernels that decrease slower than exponential (i.e. power-law
  kernels).

  We used this approach to estimate $R_0$ in wheat stripe rust (an
  important disease caused by \emph{Puccinia striiformis}), since
  disease gradients for this pathogen were thoroughly measured over
  large distances [Sackett and Mundt, Phytopathology, 95, 983
  (2005)]. For the two largest datasets, we estimated $R_0$ in the
  limit of large fields to be of the order of 50. These estimates are
  consistent with independent field observations [Cowger et
  al. (2005), Phytopathology, 95, 97282; Farber et al. (2013),
  Phytopathology, 103, 41].

  We present a proof of principle of a novel approach to estimate the
  basic reproductive number, $R_0$, of plant pathogens using wheat
  stripe rust as a case study. We found that the spatial extent over
  which $R_0$ changes strongly is quite fine-scaled (about 30 m of the
  linear extension of the field). Our results indicate that in order
  to optimize the spatial scale of deployment of fungicides or host
  resistances, the adjustments should be made at a fine spatial scale.
\end{abstract}


\section{Introduction}

When plant pathogens succeed in infecting their hosts, they colonize
the host tissue and deprive hosts of resources and energy. This often leads
to serious yield losses in agriculture
\citep{stsc05}. Disease-resistant crop varieties and chemicals
(fungicides or antibiotics) are widely used to control infectious
diseases of plants. But both of these control measures are highly
vulnerable to pathogen adaptation: pathogens evolve to overcome host
resistances and to become insensitive to fungicides \citep{mcli02}. In order to
devise effective and durable strategies of disease control \citep{mu14}, a thorough
understanding of basic epidemiological properties of plant pathogens
with the help of appropriate mathematical models is necessary.

The spread of infectious diseases depends on the contact structure,
a network in which each host is a node and has a number of weighted,
directional links to other hosts. The strength of each link represents
the probability of transmission from one host to another. In
infectious diseases of humans and animals contact structures are
determined by networks of social contacts. 
Plant pathogens spread over global scales of countries and continents
by natural means and through networks of trade and exchange
\citep{brho02,shpa14}. However, at a local scale of a single field of
crop plants or several adjacent fields, plant pathogens spread
primarily through passive dispersal of infectious propagules through
air, water or soil between immobile plants. Insect pests may disperse
both actively and passively between hosts plants \citep{mado12}. In
both of these cases, the probability of transmission between hosts
depends on the geographical distance between them. Hence, the contact
structure is determined by the spatial scales of pathogen dispersal
and the spatial scales of the host population.


Full information on the contact structure is difficult to obtain and
to analyze. Several global measures are used to characterize networks
of contacts, such as the average degree, i.\,e. the average number of
links per host. Yet, a better measure that characterizes the disease
spread is its basic reproductive number, $R_0$, defined intuitively as
``the average number of secondary cases of infection generated by one
primary case in a susceptible host population''
\citep{anma86}. Mathematically, it is given by the dominant eigenvalue
of the next generation operator \citep{he02}. 
Hence, the basic reproductive number is a quantity with a clear biological
meaning that characterizes reproductive fitness
of the pathogen. It determines the invasion threshold: if $R_0>1$ the disease
will spread in the population, otherwise at $R_0<1$ the pathogen will
eventually die out. Therefore, $R_0$ can be used to estimate the
critical proportion of the host population that needs to be immunized
(i.\,e. vaccinated) in order to eradicate the disease \citep{anma91}. Also, $R_0$
often allows one to estimate the final (equilibrium) disease level.
%

Much attention has been devoted to estimation of $R_0$ for infectious
diseases of humans and animals \citep{anma91,frdo+09,hadu+09}. Several
studies discuss $R_0$ in the context of infectious diseases of plants
\citep{gugi+00,pagu+01,pagi+05,bomc+08}, but only one study provided actual
estimates based on measurements of the apparent infection rate $r$
(the rate of growth of the disease proportion over time, assuming
logistic growth \citep{va63}) for wheat stripe rust \citep{seje+01}.
Another approach is to estimate $R_0$ by fitting the solution
of a population dynamics model of disease spread to an empirical
disease progress curve (i.\,e. the plot of the proportion of disease
over time).
%
%
However, this appears to be difficult, because we expect $R_0$ to
depend on the spatial scales of the host
population. 
%
%
In an agricultural setting, crop plants are usually arranged in nearly
rectangular fields. Each field is characterized by its area $S$ and
aspect ratio $\alpha$. Hence, $R_0$ should depend on $S$ and $\alpha$,
provided that the planting density is fixed. Given the wide variation
in field sizes and shapes across individual fields and growing
regions, countries and continents, a useful estimate for $R_0$ should
also capture the dependence on the field size and shape. But measuring
disease progress curves in many fields with different sizes and shapes
requires enormous efforts and resources.

 
In this study we propose a novel way to estimate the basic
reproductive number $R_0$ as a function of field size and shape. This
approach uses a spatially explicit population dynamics model
formulated as a system of integro-differential equations. The
estimation of $R_0$ is based on disease gradient measurements in which
the amount of disease is characterized as a function of the distance
from a localized source of initial
inoculum. 
The advantage of this approach is that, by measuring the disease gradient
over a large enough distance in a single experiment, one captures the
information on the dependence of $R_0$ on the field size and aspect
ratio. In this way, more useful information can be extracted from
disease gradient data than thought previously.

To provide a proof of principle for this method, we applied it to
wheat stripe rust (an important pathogen of wheat caused by
\emph{Puccinia striiformis} \citep{we11}), since disease gradients for this
pathogen were thoroughly measured over large distances
\citep{samu05,cowa+05}. Using these data, we estimated $R_0$ as a
function of the field size and shape. From this dependence we
determined the ranges of field sizes and shapes over which $R_0$
exhibits a considerable change.


\section{Materials and methods}
\label{sec:methods}

We assume that the hosts are continuously distributed across the
rectangular field with the dimensions $d_x$ and $d_y$. The field area
is $S = d_x d_y$ and its aspect ratio is $\alpha = d_x/d_y$, so that
$\alpha$ close to zero refers to long, narrow fields, while $\alpha=1$
represents a square field.
We trace the densities of healthy hosts $H(x,y,t)$ and infected hosts
$I(x,y,t)$ in space and time using the system of integro-differential equations
%
\begin{linenomath}
\begin{align}
  &\frac{\pd H(x, y, t)}{\pd t} = r_H H(x, y, t) \left[1 - H(x, y, t)/K
  \right] - \beta \lambda (x,y) H(x, y, t), \label{eq:host-par-ide-2d-1} \\
  &\frac{\pd I(x, y, t)}{\pd t} = \beta \lambda (x,y) H(x, y, t) - \mu I(x, y, t). \label{eq:host-par-ide-2d-2}
\end{align}
\end{linenomath}
Here, the force of infection $\lambda(x,y)$ at a location $x$, $y$ is
determined by integrating over all possible sources of infection:
\begin{linenomath}
\begin{align}\label{eq:force-inf}
\lambda = \int_0^{d_x} du \int_0^{d_y} dv \,  \kappa (x,y,u,v) I(u, v, t).
\end{align}
\end{linenomath}
In obtaining
Eqs.\,(\ref{eq:host-par-ide-2d-1})-(\ref{eq:host-par-ide-2d-2}) we
assumed that the characteristic time scale of spore dispersal is much
shorter than the characteristic time scales associated with other
stages of the pathogen life cycle and, hence, the density of spores is
proportional to the density of the infectious host tissue (see
Appendix\,A.4 in Supporting Information for more
details).

The quantities $H(x,y,t)$ and $I(x,y,t)$ represent the areas of the
corresponding host tissue per unit land area. The host tissue could be
leaves, stems or grain, depending on the specific host-pathogen
interaction.
Healthy hosts $H(x,y,t)$ grow logistically with the rate $r_H$ and the
``carrying capacity'' $K$, which may imply limited space or nutrients.
Furthermore, healthy hosts may be infected by the pathogen and
transformed into infected hosts with the rate $\beta \lambda(x,y)$. The
transmission rate $\beta$ is a compound parameter given by the product of
the sporulation rate of the infected tissue and the probability that a
spore causes new infection. Infected host tissue loses its infectivity at a rate
$\mu$, where $\mu^{-1}$ is the average infectious period.
An approximate version of the model
Eqs.\,(\ref{eq:host-par-ide-2d-1})-(\ref{eq:host-par-ide-2d-2}), in
which the host densities were assumed to be homogeneous in space, was
used in several previous studies of plant disease epidemics \citep{hagu+07,bogi08,mimc+14}.

The integral in \eq{eq:force-inf} is weighted using $\kappa(x,y,u,v)$,
the dispersal kernel (or contact distribution \citep{mo77}) that
characterizes the dispersal properties of the pathogen. The dispersal
properties as well as the environmental conditions are assumed to be
the same along the field. Moreover, dispersal is assumed to be
isotropic, meaning that a spore has the same probability to move in
any direction along the two-dimensional field. The latter assumption
can be problematic when strong winds prevail in a certain direction
and may be the cause of discrepancy with the empirical findings (see
Appendix\,A.5).
In this case, the dispersal kernel is only determined by the distance
between the source and the target of infection,
i.\,e. $\kappa(x,y,u,v) = \kappa(r)$, where $r = \sqrt{(x-u)^2 + (y -
  v)^2}$. For aerially dispersed plant diseases, $\kappa(r)$ is
defined as a probability density function for an infectious spore to
land at a distance $r$ from its source \citep{nakl+12}.

In order to determine the basic reproductive number, $R_0$, we perform
the linear stability analysis of the disease-free equilibrium of the
system Eqs.\,(\ref{eq:host-par-ide-2d-1})-(\ref{eq:host-par-ide-2d-2}). This
leads to the eigenvalue problem for the Fredholm equation of the
second kind  (see Appendix\,A.1 for the derivation)
\begin{equation}\label{eq:fredh2eigval}
R_{0\infty} \int_0^{d_x} du\, \int_0^{d_y} dv \, \kappa (r) w(u,v) = \sigma w(x,y),
\end{equation}
where $R_{0\infty} = \beta K/\mu$.  By solving this problem, we can
find the eigenvalues $\sigma_i$ and eigenfunctions $w_i(x,y)$ that
satisfy the \eq{eq:fredh2eigval}. The dominant eigenvalue $\sigma_d$
determines the basic reproductive number, i.\,e. $R_0 =
\sigma_d$. Although an approximate expression for $R_0$ based on its intuitive definition may
often give sound results, this cannot be guaranteed (see Appendix\,A.2).
%
%


\section{Results}

We first consider the generic features of how the basic reproductive
number, $R_0$, depends on the field size $d$. Then, we consider these
dependencies in the case of wheat stripe rust in
\sec{sec:case-pstriiformis}.

\subsection{Dependence of the basic reproductive number on the field size}
\label{sec:r0-vs-a}

The basic reproductive number, $R_0$, is shown in \fig{fig:r0-vs-a} as
a function of the linear extension $d$ of a square field for three
different dispersal kernels (Gaussian, exponential and modified
power-law). These three functional forms are often used to describe
dispersal gradients in plant diseases \citep{figr+87,frbo00,samu05},
but also in other taxonomic groups, for example, in pollen, seeds,
seedlings, beetles, moths and butterflies \citep{nakl+12}.
These three functions represent the three classes of dispersal
kernels: ``thin-tailed'' (Gaussian) that decrease faster than
exponential, exponential, and ``fat-tailed'' that decrease slower than
exponential (power-law). ``Thin-tailed'' and exponential kernels give
rise to travelling epidemic waves with a constant velocity, while the
``fat-tailed'' kernels result in accelerating epidemic waves
\citep{mo77,meko03,cowa+05,samu05a}.

For all the three types of dispersal kernels that we considered, the
basic reproductive number first increases as a function of the field
size $d$ and then, eventually, saturates to a constant value (\fig{fig:r0-vs-a}).
Thus, we find that the qualitative dependence of $R_0$, a more
basic epididemiological parameter than the epidemic velocity, on the
field size is quite robust with respect to the functional
form of the dispersal kernel. In particular, it is not affected much
by the nature of the tails of the dispersal kernel. Moreover, we
expect this behaviour to hold for any dispersal kernel, as long as it
a monotonically decreasing function of the distance $r$.


The initial growth of $R_0$ versus $d$ follows a quadratic function
(see Eq.\,(A.10)). It occurs because in this range, the
field size is much smaller than the dispersal radius $a$ (a characteristic
length scale of pathogen dispersal), i.\,e. $d \ll
a$. Therefore, by making the field larger, more spores will land
within the field and lead to new infections. In other words, in this
range the field size is the limiting factor for the pathogen fitness.
On the contrary, when the field size is much larger than the dispersal
radius, i.\,e. $d \gg a$, the basic reproductive number becomes
independent of $d$. Here, pathogen does not become fitter on a larger
field, because its fitness is now limited by the range of dispersal
and not by the size of the field.


While the three curves in \fig{fig:r0-vs-a} exhibit a universal
qualitative behaviour, they differ in the rate at which the saturation
occurs at large field sizes.
The Gaussian dispersal kernel decreases faster with the distance
$r$ than the exponential dispersal kernel. As a result, $R_0$ grows
and saturates as a function of the field size $d$ faster for the Gaussian
than for the exponential. 
The result for the power-law dispersal kernel is difficult to compare
with the results for other kernels, since the power law lacks a
meaningful characteristic length scale. Asymptotically, at large field
sizes $R_0$ approaches the constant value slower in the case of the
power-law dispersal kernel than for the other two kernels. However,
at small field sizes, $R_0$ as a function of $d$ may grow faster or
slower for the power-law kernel as compared to the other two kernels,
depending on the values of the parameters $r_0$ and $b$.
In \fig{fig:r0-vs-a}, we present an example when the $R_0$ for the
power law first grows faster than the that for the Gaussian or
exponential dispersal kernels, but subsequently its growth slows down
and becomes slower than for the Gaussian and exponential (as expected
from the asymptotic behavior of the corresponding dispersal kernels).

\subsection{Case study: dependence of the basic reproductive number on
  the field size and shape for wheat stripe rust}
\label{sec:case-pstriiformis}

We infer the dependence of the basic reproductive number, $R_0$, on
the field size and shape from the detailed measurements of primary
disease gradients of wheat stripe rust \citep{samu05,cowa+05}. $R_0$
is computed by numerically solving the eigenvalue problem in
\eq{eq:fredh2eigval} for different values of the field dimensions
$d_x$ and $d_y$ that characterize the field size and shape. To perform
this calculation, we estimated the dispersal kernel $\kappa(r)$ and the
compound parameter $R_{0\infty}$ that corresponds to the basic
reproductive number for a very large field from experimental data
\citep{samu05,cowa+05} [see Appendix\,A.3 for the
details of the estimation procedure].

In these experiments, small areas of experimental plots (foci) were
artificially inoculated by pathogen spores ($0^\mrm{th}$
generation). These spores give rise to lesions in the focus (first
generation) that further produce spores, which are dispersed through
the air. This gives rise to infection outside of the focus, producing
the second generation of pathogen lesions. The corresponding disease
severity (the proportion of the leaf area infected) is measured as a
function of the distance $r$ from the focus.

The outcome of this measurement is shown in
\fig{fig:disgrad-data-fit-herm02} for the two largest datasets
(Hermiston 2002 and Madras 2002, downwind) obtained in this
experiment.  These two datasets were chosen because they contained
measurements over large enough distances that allowed us to obtain
sound fits. Disease severity strongly depends on the distance $r$: the
value is largest closer to the focus and decreases monotonically with
$r$.  The data can be fitted well by the modified power-law function
(solid curve in \fig{fig:disgrad-data-fit-herm02})
\begin{equation}\label{eq:mpowlaw-kern2}
\kappa_\mrm{PL2}(r) = \kappa_0 \left(r_0^2 + r^2 \right)^{-b/2}.
\end{equation} 
In contrast, exponential and Gaussian functions provide poor fits
(dashed and dotted curves in \fig{fig:disgrad-data-fit-herm02}).
(For more details on fitting see Appendix\,A.3.1 and Fig.\,6 in the
Electronic Supplementary Materials).

Disease gradients, measured in this way, contain information on
the three key processes in the pathogen life-cycle: spore production,
aerial movement of spores, and infection of healthy host tissue. We assume
that the rate of spore production and the probability to infect
healthy host tissue, once the spore has landed on it, are homogeneous
across the field, i.\,e. do not depend on the distance $r$. Hence, the
compound parameter $R_{0\infty} =\beta K / \mu$ that characterizes
these processes does not depend on the distance.  Therefore, the aerial
movement of spores is the only process that depends on the distance
$r$.
%
Further, we assume that there is a large enough number
of spores produced and the probability of infection is large enough
such that the recorded disease severity is proportional to the spore
concentration in the air. Under these assumptions, our estimate for
the dispersal kernel $\kappa(r)$ is the modified power-law function
[\eq{eq:mpowlaw-kern2}] fitted to the disease gradient data and
normalized as a probability density function (i.\,e. such that its
integral over the whole two-dimensional space equals to unity
[Appendix\,A.3.2]).
We also estimated the parameter $R_{0\infty} $ from the disease
gradient data (see Appendix\,A.3.3) and obtained the value
$R_{0\infty} =65.0$ for the Hermiston 2002 downwind dataset; and the
value $R_{0\infty} =38.0$ for the Madras 2002 downwind dataset.





Using our estimates for the dispersal kernel, $\kappa(r)$, and the
parameter $R_{0\infty} $ we solved the eigenvalue problem
\eq{eq:fredh2eigval} numerically for different field sizes and
shapes. In this way, we obtained the dependence of the basic
reproductive number $R_0$ on the field size
(\fig{fig:r0-vs-a-empir-dprogr}) and its aspect ratio
(\fig{fig:r0-vs-arat-herm02dw}).
%
%
%
In \fig{fig:r0-vs-a-empir-dprogr}, $R_0$ first grows steeply versus
the linear extension of a square field and saturates towards the
asymptotic value $R_{0\infty}$ for large fields. The basic
reproductive number is about two times larger for the parameter values
corresponding to Hermiston 2002 dataset, than for the case of Madras
2002 dataset. This difference stems from the difference in the
asymptotic values $R_{0\infty}$ and also from different shapes of the
disease gradients (cf. panel (a) and (b) in
\fig{fig:disgrad-data-fit-herm02}). 


%
%
%

The asymptotic value, $R_{0\infty}$, (indicated by the horizontal
dashed line in \fig{fig:r0-vs-a-empir-dprogr}), is approached faster
in the case of Hermiston 2002 dataset (solid curve in
\fig{fig:r0-vs-a-empir-dprogr}), than for Madras 2002 dataset (dashed
curve in \fig{fig:r0-vs-a-empir-dprogr}). The reason for this is a
different exponent of the power-law function that best fits the
corresponding disease gradients ($b = 3.04$ for Hermiston 2002,
Eq.\,(A.15), and $b = 2.23$, Eq.\,(A.16)). The
disease gradient in Madras 2002 decreases slower due a lower
exponent.

In \fig{fig:r0-vs-arat-herm02dw}, $R_0$ exhibits a saturating growth
as the field aspect ratio $\alpha$ is increased from 0.01 to 1. Hence,
the square fields, with $\alpha=1$, are most conducive for the
disease growth.  The basic reproductuve number grows faster and
saturates at larger values of $\alpha$ in smaller fields (cf. dotted,
dashed, dash-dotted and solid curves in
\fig{fig:r0-vs-arat-herm02dw}).

A number of empirical studies have reported that, in agreement with our
results, smaller plots resulted in lower disease levels in wheat
yellow rust \citep{mubr+96}, wheat brown rust (\emph{Puccinia
  recondita} f. sp. \emph{tritici}) \citep{bote+84}, potato late
blight \citep{pafr83} and \emph{Valdensia heterodoxa} on
\emph{Vaccinium myrtillus} \citep{sten+06}. However, in a more recent
study in wheat yellow rust \citep{samu09} that used considerably
larger plot sizes, the plot size did not affect the epidemic
velocity. Our estimation framework predicts moderate differences in
the values of $R_0$ between larger square plots and smaller
elongated plots used in experiments \citep{samu09} (cf. the white and
gray circles in both panels of \fig{fig:r0-vs-arat-herm02dw}). This is
expected to result in higher epidemic velocities in larger plots
compared to smaller plots, according to theoretical arguments (see
Appendix\,A.5).
%
%
We suggest two possible explanations for this discrepancy
(see Appendix\,A.5 for more details).  First, strong wind
with a prevailing direction along the axis of the elongated plot was
observed in the experimental setting \citep{samu09}, but in our model
isotropic dispersal was assumed. The differences in $R_0$ between
smaller elongated plot and a larger square plot that we predict using
the model are possibly masked by the wind. This is because the wind
may increase the pathogen's $R_0$ in the smaller elongated plot by
preventing the spores to land outside the plot. Second, a moderate
difference of 20-30\,\% that we predict for epidemic velocities may be
difficult to detect given the level of experimental uncertainties.







\section{Discussion}
\label{sec:discussion}


We found that the basic reproductive number, $R_0$, of crop pathogens
depends on the size and geometry of the field planted with host
plants. $R_0$ increases with the field size at small field sizes and
saturates to a constant values at large field sizes. The value of
$R_0$ reaches its maximum in square fields and decreases as the
field becomes elongated, while retaining the same area.
%
This pattern appears to be quite general: it holds for dispersal
kernels that decrease exponentially or faster (i.\,e. Gaussian
kernels) as well as for ``fat-tailed'' dispersal kernels that decrease
slower than exponential ones (i.\,e. power-law kernels). We expect the
same qualitative behavior for any dispersal kernel, provided that it
is a monotonically decreasing function of the distance.




%
%
As expected, this qualitative picture also holds for the dispersal
kernels estimated in wheat stripe rust.
The asymptotic values of the basic reproductive number at large field
sizes ($R_{0\infty} =65.0$ for Hermiston 2002 downwind, $R_{0\infty}
=38.0$ for the Madras 2002 downwind dataset) result in the values of
the apparent rate of infection $r\approx 0.21$ for Hermiston and $r
\approx 0.18$ for Madras, where the simple relationship $r = \mu \log
R_0$ was used. These values are quite close to the estimates of $r$
obtained independently for these experiments ($r \approx 0.25$
\citep{cowa+05}). Also, in \citep{seje+01} the $R_0$ of wheat yellow
rust was estimated to be around 60 from the measurements of the
apparent rate of infection $r$. This study used a more rigorous
approach to connect $r$ and $R_0$ that took into account the shape of
the sporulation curve. Our estimates of $R_{0\infty}$ are also
consistent, but somewhat smaller than the estimates from field
experiments, where the number of secondary lesions originating from a
single lesion was measured to be as high as several hundred
\citep{faes+13}.



The estimates for $R_{0\infty}$ that we obtained for wheat stripe rust
are considerably larger than typical estimates for the basic
reproductive number for human or animal diseases. For example, the
relatively large values of $R_0$ were estimated for childhood diseases
such as measles (14-18) and pertussis (5-18) \citep{anma91}, the
estimates for the ``swine flu'' influenza H1N1 were in the range
1.4-1.6 \citep{frdo+09}, the estimates for rabies were in the range
1-2 \citep{hadu+09}. A possible exception is malaria, where the
estimates of $R_0$ between one and more than 3000 were reported
\citep{smmc+07}. The $R_0$ determines the critical proportion $p_c$ of
the host population that needs to be immunized in order to eradicate
the disease ($p_c = 1 - 1/R_0$) \citep{anma91}. For example, our estimate for the
wheat stripe rust of $R_0 \simeq 50$ yields the critical proportion
$p_c \simeq 0.98$. This may explain why it is so difficult to
eradicate rusts, while there are cases of dangerous human diseases
(for example, small pox) that were eradicated with
the help of vaccination programmes \citep{anma91}.
%
%
This difference in the values of $R_0$ may
result from a different biology of hosts (animals versus plants), or,
alternatively, it could be due to different nature of the diseases,
i.\,e. systemic diseases in the case of humans and animals versus
local lesion diseases in the case of wheat stripe rust.
To determine which of these two explanations is more plausible, one
needs to estimate $R_0$ for systemic disease of plants and local
lesion (i.\,e. skin diseases) of animals.
%
%
This difference may also be caused by the characteristic features of
host populations in agroecosystems, where genetically uniform hosts
are planted with high densities in a homogeneous environment. Hence,
it would be interesting to compare the $R_0$ of crop pathogens with
the $R_0$ of plant pathogens in natural ecosystems.


These findings can be used to control plant diseases, if one knows the
spatial scales, i.\,e. field sizes and aspect ratios, over which $R_0$
changes considerably.  We found that the $R_0$ of wheat stripe rust
exhibits a large change at a fine spatial scale: when the linear
dimension of a square field increases from zero to about thirty
meters (\fig{fig:r0-vs-a-empir-dprogr}). The most substantial change
of $R_0$ as a function of the field aspect ratio occurs between aspect
ratios of 0.01 and 0.2. These results suggest, that decreasing field
sizes and elongating fields may not be a practical measure to control
wheat stripe rust, because the beneficial effect of lowering the
disease levels is in this case unlikely to outweigh the economical
costs associated with using smaller and longer fields.
But this method could be feasible for controlling other diseases of
crops or pests (for example, western corn rootworm that can disperse
over longer distances \citep{caha+10} than wheat stripe rust).
We hope that our study will stimulate more detailed empirical studies
of the disease gradients for different crop pathogens over long
distances, such that the framework proposed here could be used to infer
how the $R_0$ depends on the spatial scales of the host population.
%
%
%
Although similar ideas about possibilities to control plant diseases
by adjusting field size and geometry were explored mathematically in
\citep{flma+82}, their framework based on reaction-diffusion models
was not capable of including realistic dispersal kernels. Hence, they
could not estimate the spatial scales at which the pathogen fitness
changes considerably.

The experiments in Hermiston 2002 and Madras 2002 used the same
planting density, the same wheat cultivar and the same pathogen race
was used for initial inoculation. But the environmental conditions
were somewhat different in these two locations. Hence, we can largely
attribute the difference in the disease gradients between these two
datasets and the resulting difference in the estimated values of the
basic reproductive number to the difference in the environmental
conditions. In contrast, in natural epidemics the variation in the
outcomes of pathogen dispersal can also result from the genetic
variation in pathogen and host population \citep{taha+14}. Therefore,
in would be interesting to explore the effect of simulataneously
adjusting the spatial scales and introducing genetic diversity to the
host population by using host mixtures or multiline cultivars
\citep{mu02,mimc+14a}

From another point of view, our findings could be helpful for choosing
the minimum plot sizes and aspect ratios for field experimentation in
plant pathology. For the experimental plots to be representative of larger fields
used by growers, the plot size and aspect ratio should
be chosen such that they correspond to the start of the saturation of
the dependency of $R_0$ on the field size
(\fig{fig:r0-vs-a-empir-dprogr}) and aspect ratio
(\fig{fig:r0-vs-arat-herm02dw}). Thus, our results indicate that
in the case of wheat stripe rust, the area of experimental plots
should be at least 0.25\,ha and the aspect ratio should be at least
0.2 (this corresponds approximately to a 20\,m$\times$110\,m elongated
plot, or, alternatively, a 50\,m$\times$50\,m square plot).




Our results could also help to manage fungicide resistance: if several
different fungicides are applied over smaller, elongated patches
within a larger field, then the fitness of resistant strains would be
diminished. This strategy allows one to keep the overall field size large
enough to be economically advantageous, but requires availability of
several different fungicides that have little or no cross-resistance.
The same reasoning applies also for the case of break-down of disease
resistance in host plants. In this case, host cultivars with different
disease resistances should be arranged in smaller, elongated patches
within a larger field.
Favorable arrangements of these patches with different fungicides and
host cultivars that would reduce selection for fungicide resistance
and minimize break-down of host defences can be investigated using
dynamical simulations of the population dynamics model based on
Eqs.\,(\ref{eq:host-par-ide-2d-1})-(\ref{eq:host-par-ide-2d-2}). In order to suggest economically viable implementations,
an epidemiological modeling framework should be coupled with a sound
economical cost-benefit analysis. 





So far we discussed disease control on the level of a single field of
crops. But our study also provides a way to incorporate the
dependence of $R_0$ on the spatial stucture of the local host
population into models of disease spread on a regional scale (such as
the models described in \citep{pabo+06,pato+14}). In this
context we expect the nature of tails of the dispersal kernels to play
an important role in the disease spread and would influence optimal
strategies of disease control.


\section{Acknowledgements}

AM and SB gratefully acknowledge financial support by the ERC advanced
grant PBDR 268540 ``The population biology of drug resistance: Key
principles for a more sustainable use of drugs''. The contributions of
CM were supported by NIH grant R01GM96685 through the NSF/NIH Ecology
and Evolution of Infectious Disease Program. The authors would like to
thank Kathryn Sackett for providing estimates of the apparent
infection rate and helpful discussions. AM is grateful to Bruce
McDonald and Roland Regoes for many valuable discussions.











\begin{thebibliography}{49}
\expandafter\ifx\csname natexlab\endcsname\relax\def\natexlab#1{#1}\fi
\expandafter\ifx\csname url\endcsname\relax
  \def\url#1{{\tt #1}}\fi
\expandafter\ifx\csname urlprefix\endcsname\relax\def\urlprefix{URL }\fi

\bibitem[{Anderson and May(1986)}]{anma86}
Anderson, R.~M., and R.~M. May.
\newblock 1986.
\newblock {The invasion, persistence and spread of infectious diseases within
  animal and plant communities.}
\newblock Philosophical transactions of the Royal Society of London. Series B,
  Biological sciences {\bf 314}:533--70.

\bibitem[{Anderson and May.(1991)}]{anma91}
Anderson, R.~M., and R.~M. May.
\newblock 1991.
\newblock {Infectious diseases of humans}.
\newblock Oxford University Press.

\bibitem[{Bowen et~al.(1984)Bowen, Teng, and Roelfs}]{bote+84}
Bowen, K., P.~Teng, and A.~Roelfs.
\newblock 1984.
\newblock {Negative Interplot Interference in Field Experiments with Leaf Rust
  of Wheat}.
\newblock Phytopathology {\bf 74}:1157--61.

\bibitem[{Brown and Hovmoller(2002)}]{brho02}
Brown, J. K.~M., and M.~S. Hovmoller.
\newblock 2002.
\newblock {Aerial dispersal of pathogens on the global and continental scales
  and its impact on plant disease.}
\newblock Science (New York, N.Y.) {\bf 297}:537--41.

\bibitem[{Carrasco et~al.(2010)Carrasco, Harwood, Toepfer, MacLeod, Levay,
  Kiss, Baker, Mumford, and Knight}]{caha+10}
Carrasco, L., T.~Harwood, S.~Toepfer, A.~MacLeod, N.~Levay, J.~Kiss, R.~Baker,
  J.~Mumford, and J.~Knight.
\newblock 2010.
\newblock {Dispersal kernels of the invasive alien western corn rootworm and
  the effectiveness of buffer zones in eradication programmes in Europe}.
\newblock Annals of Applied Biology {\bf 156}:63--77.

\bibitem[{Cowger et~al.(2005)Cowger, Wallace, and Mundt}]{cowa+05}
Cowger, C., L.~D. Wallace, and C.~C. Mundt.
\newblock 2005.
\newblock {Velocity of spread of wheat stripe rust epidemics.}
\newblock Phytopathology {\bf 95}:972--82.

\bibitem[{Diekmann et~al.(1990)Diekmann, Heesterbeek, and Metz}]{dihe+90}
Diekmann, O., J.~A.~P. Heesterbeek, and J.~A.~J. Metz.
\newblock 1990.
\newblock {On the definition and the computation of the basic reproduction
  ratio R0 in models for infectious diseases in heterogeneous populations}.
\newblock Journal of Mathematical Biology {\bf 28}:365--382.

\bibitem[{Farber et~al.(2013)Farber, Estep, Sackett, and Mundt}]{faes+13}
Farber, D., L.~Estep, K.~Sackett, and C.~Mundt.
\newblock 2013.
\newblock {Local dispersal of Puccinia striiformis f.sp. triticiti from single
  source lesions}.
\newblock Phytopathology APS-MSA Joint Meeting, Austin TX {\bf 103}:41.

\bibitem[{Fitt et~al.(1987)Fitt, Gregory, Todd, McCartney, and
  Macdonald}]{figr+87}
Fitt, B. D.~L., P.~H. Gregory, a.~D. Todd, H.~a. McCartney, and O.~C.
  Macdonald.
\newblock 1987.
\newblock {Spore Dispersal and Plant Disease Gradients; a Comparison between
  two Empirical Models}.
\newblock Journal of Phytopathology {\bf 118}:227--242.

\bibitem[{Fleming et~al.(1982)Fleming, Marsh, and Tuckwell}]{flma+82}
Fleming, R., L.~Marsh, and H.~Tuckwell.
\newblock 1982.
\newblock {Effect of field geometry on the spread of crop disease}.
\newblock Protection Ecology {\bf 4}:81.

\bibitem[{Frantzen and Bosch(2000)}]{frbo00}
Frantzen, J., and F.~V.~D. Bosch.
\newblock 2000.
\newblock {Spread of organisms: can travelling and dispersive waves be
  distinguished ?}
\newblock Basic and Applied Ecology {\bf 91}:83--91.

\bibitem[{Fraser et~al.(2009)Fraser, Donnelly, Cauchemez, Hanage, {Van
  Kerkhove}, Hollingsworth, Griffin, Baggaley, Jenkins, Lyons, Jombart,
  Hinsley, Grassly, Balloux, Ghani, Ferguson, Rambaut, Pybus, Lopez-Gatell,
  Alpuche-Aranda, Chapela, Zavala, Guevara, Checchi, Garcia, Hugonnet, and
  Roth}]{frdo+09}
Fraser, C., C.~a. Donnelly, S.~Cauchemez, W.~P. Hanage, M.~D. {Van Kerkhove},
  T.~D. Hollingsworth, J.~Griffin, R.~F. Baggaley, H.~E. Jenkins, E.~J. Lyons,
  T.~Jombart, W.~R. Hinsley, N.~C. Grassly, F.~Balloux, A.~C. Ghani, N.~M.
  Ferguson, A.~Rambaut, O.~G. Pybus, H.~Lopez-Gatell, C.~M. Alpuche-Aranda,
  I.~B. Chapela, E.~P. Zavala, D.~M.~E. Guevara, F.~Checchi, E.~Garcia,
  S.~Hugonnet, and C.~Roth.
\newblock 2009.
\newblock {Pandemic potential of a strain of influenza A (H1N1): early
  findings.}
\newblock Science (New York, N.Y.) {\bf 324}:1557--61.

\bibitem[{Gregory(1968)}]{gr68}
Gregory, P.
\newblock 1968.
\newblock {Interpreting plant disease dispersal gradients}.
\newblock Annual Review of Phytopathology {\bf 6}:189.

\bibitem[{Gubbins et~al.(2000)Gubbins, Gilligan, and Kleczkowski}]{gugi+00}
Gubbins, S., C.~A. Gilligan, and A.~Kleczkowski.
\newblock 2000.
\newblock {Population dynamics of plant-parasite interactions: thresholds for
  invasion.}
\newblock Theoretical Population Biology {\bf 57}:219--33.

\bibitem[{Hall et~al.(2007)Hall, Gubbins, and Gilligan}]{hagu+07}
Hall, R.~J., S.~Gubbins, and C.~A. Gilligan.
\newblock 2007.
\newblock {Evaluating the performance of chemical control in the presence of
  resistant pathogens.}
\newblock Bulletin of mathematical biology {\bf 69}:525--37.

\bibitem[{Hampson et~al.(2009)Hampson, Dushoff, Cleaveland, Haydon, Kaare,
  Packer, and Dobson}]{hadu+09}
Hampson, K., J.~Dushoff, S.~Cleaveland, D.~T. Haydon, M.~Kaare, C.~Packer, and
  A.~Dobson.
\newblock 2009.
\newblock {Transmission dynamics and prospects for the elimination of canine
  rabies.}
\newblock PLoS Biology {\bf 7}:e53.

\bibitem[{Heesterbeek(2002)}]{he02}
Heesterbeek, J.
\newblock 2002.
\newblock {A Brief History of R 0 and a Recipe for its Calculation}.
\newblock Acta Biotheoretica {\bf 5}:189--204.

\bibitem[{Keeling and Rohani(2008)}]{kero08epvel}
Keeling, M.~J., and P.~Rohani, 2008.
\newblock {Spatial Models}.
\newblock Chapter~7, page 266 {\em in\/} Modeling Infectious Diseases in Humans
  and Animals. Princeton University Press, Princeton.

\bibitem[{Lambert et~al.(1980)Lambert, Villareal, Mackenzie, and
  Agri}]{lavi+80}
Lambert, D.~H., R.~L. Villareal, D.~R. Mackenzie, and T.~P. Agri.
\newblock 1980.
\newblock {A General Model for Gradient Analysis}.
\newblock Phytopathol. Z. {\bf 154}:150--155.

\bibitem[{Mazzi and Dorn(2012)}]{mado12}
Mazzi, D., and S.~Dorn.
\newblock 2012.
\newblock {Movement of insect pests in agricultural landscapes}.
\newblock Annals of Applied Biology {\bf 160}:97--113.

\bibitem[{McDonald and Linde(2002)}]{mcli02}
McDonald, B.~A., and C.~Linde.
\newblock 2002.
\newblock {Pathogen population genetics, evolutionary potential, and durable
  resistance.}
\newblock Annual Review of Phytopathology {\bf 40}:349--79.

\bibitem[{Medlock and Kot(2003)}]{meko03}
Medlock, J., and M.~Kot.
\newblock 2003.
\newblock {Spreading disease: integro-differential equations old and new}.
\newblock Mathematical Biosciences {\bf 184}:201--222.

\bibitem[{Mikaberidze et~al.(2014{\natexlab{{\em a\/}}})Mikaberidze, McDonald,
  and Bonhoeffer}]{mimc+14a}
Mikaberidze, A., B.~McDonald, and S.~Bonhoeffer.
\newblock 2014{\natexlab{{\em a\/}}}.
\newblock {How to develop smarter host mixtures to control plant disease?}
\newblock arXiv:1402.2788 .

\bibitem[{Mikaberidze et~al.(2014{\natexlab{{\em b\/}}})Mikaberidze, McDonald,
  and Bonhoeffer}]{mimc+14}
Mikaberidze, A., B.~A. McDonald, and S.~Bonhoeffer.
\newblock 2014{\natexlab{{\em b\/}}}.
\newblock {Can high risk fungicides be used in mixtures without selecting for
  fungicide resistance?}
\newblock Phytopathology {\bf 104}:324--331.

\bibitem[{Mollison(1977)}]{mo77}
Mollison, D.
\newblock 1977.
\newblock {Spatial Contact Models for Ecological and Epidemic Spread}.
\newblock Journal of the Royal Statistical Society. Series B {\bf 39}:283.

\bibitem[{Mundt and Leonard(1985)}]{mule85}
Mundt, C., and K.~Leonard.
\newblock 1985.
\newblock {A modification of Gregory's model for describing plant disease
  gradients.}
\newblock Phytopathology {\bf 75}:930.

\bibitem[{Mundt(2002)}]{mu02}
Mundt, C.~C.
\newblock 2002.
\newblock {Use of multiline cultivars and cultivar mixtures for disease
  management.}
\newblock Annual review of phytopathology {\bf 40}:381--410.

\bibitem[{Mundt(2014)}]{mu14}
Mundt, C.~C.
\newblock 2014.
\newblock {Durable resistance: A key to sustainable management of pathogens and
  pests.}
\newblock Infection, genetics and evolution: journal of molecular epidemiology
  and evolutionary genetics in infectious diseases {\bf in press}.

\bibitem[{Mundt et~al.(1996)Mundt, Brophy, and Kolar}]{mubr+96}
Mundt, C.~C., L.~S. Brophy, and S.~C. Kolar.
\newblock 1996.
\newblock {Effect of genotype unit number and spatial arrangement on severity
  of yellow rust in wheat cultivar mixtures}.
\newblock Plant Pathology {\bf 45}:215--222.

\bibitem[{Nathan et~al.(2012)Nathan, Klein, Robledo-Arnuncio, and
  Revilla}]{nakl+12}
Nathan, R., E.~Klein, J.~J. Robledo-Arnuncio, and E.~Revilla, 2012.
\newblock {Dispersal kernels: review}.
\newblock Chapter~15, page 187 {\em in\/} J.~Clobert, M.~Baguette, T.~G.
  Benton, and J.~M. Bullock, editors. Dispersal ecology and evolution. Oxford
  Univ. Press.

\bibitem[{Papa\"{\i}x et~al.(2014)Papa\"{\i}x, Touzeau, Monod, and
  Lannou}]{pato+14}
Papa\"{\i}x, J., S.~Touzeau, H.~Monod, and C.~Lannou.
\newblock 2014.
\newblock {Can epidemic control be achieved by altering landscape connectivity
  in agricultural systems?}
\newblock Ecological Modelling {\bf 284}:35--47.

\bibitem[{Park et~al.(2001)Park, Gubbins, and Gilligan}]{pagu+01}
Park, A.~W., S.~Gubbins, and C.~A. Gilligan.
\newblock 2001.
\newblock {Invasion and persistence of plant parasites in a spatially
  structured host population}.
\newblock Oikos {\bf 94}:162--174.

\bibitem[{Parnell et~al.(2005)Parnell, Gilligan, and van~den Bosch}]{pagi+05}
Parnell, S., C.~a. Gilligan, and F.~van~den Bosch.
\newblock 2005.
\newblock {Small-scale fungicide spray heterogeneity and the coexistence of
  resistant and sensitive pathogen strains.}
\newblock Phytopathology {\bf 95}:632--9.

\bibitem[{Parnell et~al.(2006)Parnell, van~den Bosch, and Gilligan}]{pabo+06}
Parnell, S., F.~van~den Bosch, and C.~a. Gilligan.
\newblock 2006.
\newblock {Large-scale fungicide spray heterogeneity and the regional spread of
  resistant pathogen strains.}
\newblock Phytopathology {\bf 96}:549--55.

\bibitem[{Paysour and Fry(1983)}]{pafr83}
Paysour, R., and W.~Fry.
\newblock 1983.
\newblock {Interplot interference: A model for planning field experiments with
  aerially disseminated pathogens}.
\newblock Phytopathology {\bf 73}:1014.

\bibitem[{Press et~al.(1992)Press, Teukolsky, Vetterling, and
  Flannery}]{prte+92}
Press, W.~H., S.~A. Teukolsky, W.~T. Vetterling, and B.~P. Flannery, 1992.
\newblock {Chapter 18. Integral Equations and Inverse Theory}.
\newblock Page 788 {\em in\/} Numerical Recipes in C. Cambridge University
  Press, Cambridge.

\bibitem[{Sackett and Mundt(2005{\natexlab{{\em a\/}}})}]{samu05}
Sackett, K.~E., and C.~C. Mundt.
\newblock 2005{\natexlab{{\em a\/}}}.
\newblock {Primary disease gradients of wheat stripe rust in large field
  plots.}
\newblock Phytopathology {\bf 95}:983--91.

\bibitem[{Sackett and Mundt(2005{\natexlab{{\em b\/}}})}]{samu05a}
Sackett, K.~E., and C.~C. Mundt.
\newblock 2005{\natexlab{{\em b\/}}}.
\newblock {The effects of dispersal gradient and pathogen life cycle components
  on epidemic velocity in computer simulations.}
\newblock Phytopathology {\bf 95}:992--1000.

\bibitem[{Sackett and Mundt(2009)}]{samu09}
Sackett, K.~E., and C.~C. Mundt.
\newblock 2009.
\newblock {Effect of plot geometry on epidemic velocity of wheat yellow rust}.
\newblock Plant Pathology {\bf 58}:370--377.

\bibitem[{Segarra et~al.(2001)Segarra, Jeger, and van~den Bosch}]{seje+01}
Segarra, J., M.~J. Jeger, and F.~van~den Bosch.
\newblock 2001.
\newblock {Epidemic dynamics and patterns of plant diseases.}
\newblock Phytopathology {\bf 91}:1001--10.

\bibitem[{Shaw and Pautasso(2014)}]{shpa14}
Shaw, M.~W., and M.~Pautasso.
\newblock 2014.
\newblock {Networks and Plant Disease Management: Concepts and Applications}.
\newblock Annual Review of Phytopathology {\bf 52}:in press.

\bibitem[{Smith et~al.(2007)Smith, McKenzie, Snow, and Hay}]{smmc+07}
Smith, D.~L., F.~E. McKenzie, R.~W. Snow, and S.~I. Hay.
\newblock 2007.
\newblock {Revisiting the basic reproductive number for malaria and its
  implications for malaria control.}
\newblock PLoS Biology {\bf 5}:e42.

\bibitem[{Strange and Scott(2005)}]{stsc05}
Strange, R.~N., and P.~R. Scott.
\newblock 2005.
\newblock {Plant disease: a threat to global food security.}
\newblock Annual Review of Phytopathology {\bf 43}:83--116.

\bibitem[{Strengbom et~al.(2006)Strengbom, Englund, and Ericson}]{sten+06}
Strengbom, J., G.~Englund, and L.~Ericson.
\newblock 2006.
\newblock {Experimental scale and precipitation modify effects of nitrogen
  addition on a plant pathogen}.
\newblock Journal of Ecology {\bf 94}:227--233.

\bibitem[{Tack et~al.(2013)Tack, Hakala, Pet\"{a}j\"{a}, Kulmala, and
  Laine}]{taha+14}
Tack, A., J.~Hakala, T.~Pet\"{a}j\"{a}, M.~Kulmala, and A.~Laine.
\newblock 2013.
\newblock {Genotype and spatial structure shape pathogen dispersal and disease
  dynamics at small spatial scales}.
\newblock Ecology {\bf 95}:703--714.

\bibitem[{van~den Bosch and Gilligan(2008)}]{bogi08}
van~den Bosch, F., and C.~A. Gilligan.
\newblock 2008.
\newblock {Models of fungicide resistance dynamics}.
\newblock Annu. Rev. Phytopathol. {\bf 46}:123--47.

\bibitem[{van~den Bosch et~al.(2008)van~den Bosch, McRoberts, van~den Berg, and
  Madden}]{bomc+08}
van~den Bosch, F., N.~McRoberts, F.~van~den Berg, and L.~Madden.
\newblock 2008.
\newblock {The Basic Reproduction Number of Plant Pathogens: Matrix Approaches
  to Complex Dynamics}.
\newblock Phytopathology {\bf 98}:239--249.

\bibitem[{Vanderplank(1963)}]{va63}
Vanderplank, J.~E.
\newblock 1963.
\newblock {Plant Diseases: Epidemics and Control}.
\newblock Academic Press, New York.

\bibitem[{Wellings(2011)}]{we11}
Wellings, C.~R.
\newblock 2011.
\newblock {Global status of stripe rust: a review of historical and current
  threats}.
\newblock Euphytica {\bf 179}:129--141.

\end{thebibliography}

\clearpage

\begin{table}\caption{Variables and parameters}\label{tab:vars}
\begin{tabular}{ l l c }
\hline
   & Description & Dimension \\
\hline
  Variables &  &  \\
  $H(x,y,t)$ & Density of healthy host tissue & dl \\
 $I(x,y,t)$ & Density of infected host tissue & dl \\
 Parameters &  &  \\
 $d_x$, $d_y$ & Linear dimensions of the field along $x$ and $y$ & m\\
 $a$ & Characteristic spatial scale of pathogen dispersal (dispersal radius) & m\\
 $\beta$ & Transmission rate & days$^{-1}$ \\
$\mu^{-1}$ & Average infectious period & days\\
$r_H$ & Growth rate of healthy host tissue & days$^{-1}$\\
$K$ & ``Carrying capacity'' of the healthy host tissue & dl \\
$R_{0\infty}$ & Basic reproductive number in the limit of a very large field & dl\\
 Functions &  &  \\
 $\kappa(r)$ & Dispersal kernel & m$^{-1}$ \\
 $R_0(d_x, d_y)$ & Basic reproductive number & dl \\
$\lambda(x, y)$ & The force of infection [\eq{eq:force-inf}]\\
\hline
\end{tabular}
\end{table}

\clearpage

\begin{figure}
  \centerline{\includegraphics[width=0.8\textwidth]{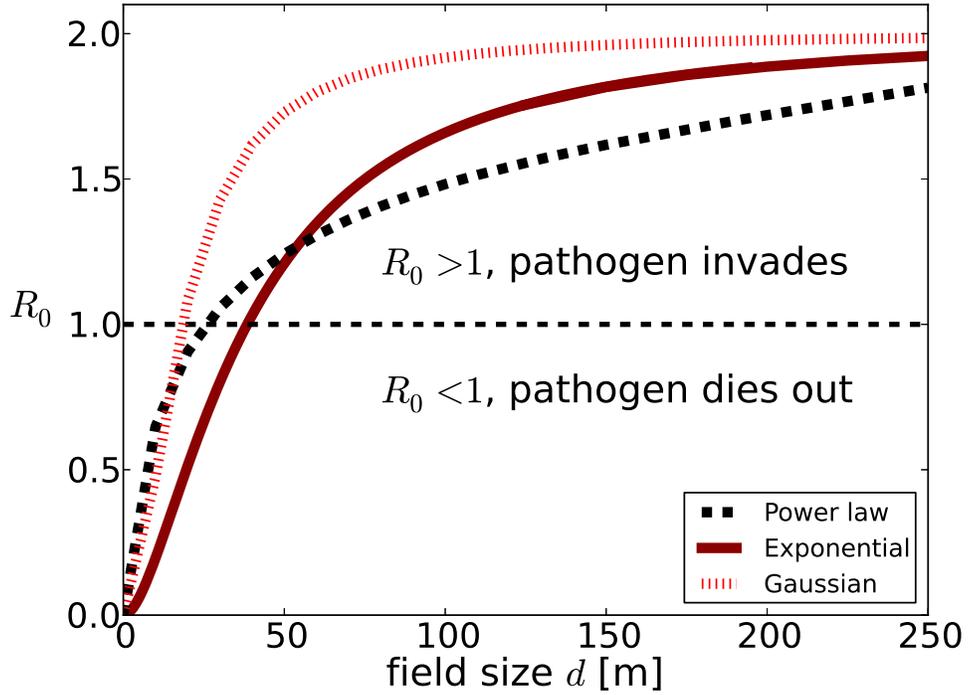}} \caption{
    Basic reproductive number $R_0$ as a function of the field size
    $d$ for the     two-dimensional field according to the numerical solution of
    \eq{eq:fredh2eigval} (solid green) using (i) the Gaussian
    [Eq.\,(A.21) at $n=2$, $a=10$\,m], (ii) the exponential
    [Eq.\,(A.21) at $n=1$, $a=10$\,m] and (iii) the power law
    dispersal kernel [Eq.\,(A.19) at $r_0=1$\, m,
    $b=2.1$]. Model parameters: $R_{0\infty}= \beta K / \mu =
    2$.}
\label{fig:r0-vs-a}
\end{figure}%
%

\clearpage
\newpage

\begin{figure}
  \centerline{\includegraphics[width=0.8\textwidth]{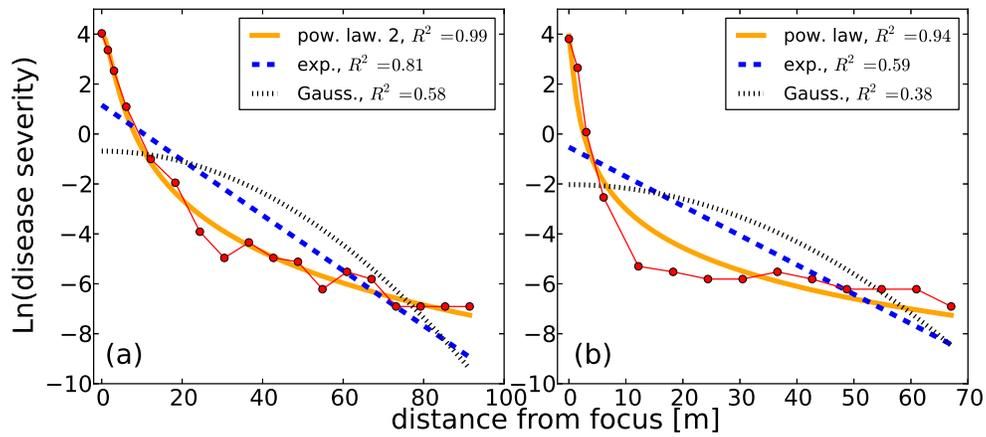}} \caption{Disease
    severity of wheat stripe rust is plotted as a function of the
    distance from focus, outcome of field experiments
    \citep{samu05,cowa+05}.  Two datasets, Hermiston 2002 downwind
    (left panel) and Madras 2002 downwind were fitted with the
    exponential function [Eq.\,(A.21) with $n=1$, dashed
    curve], the Gaussian function [Eq.\,(A.21) with $n=2$,
    dotted curve] and the modified power-law function
    [Eq.\,(A.19), solid curve].}
\label{fig:disgrad-data-fit-herm02}
\end{figure}%
%

\newpage

\begin{figure}
  \centerline{\includegraphics[width=0.8\textwidth]{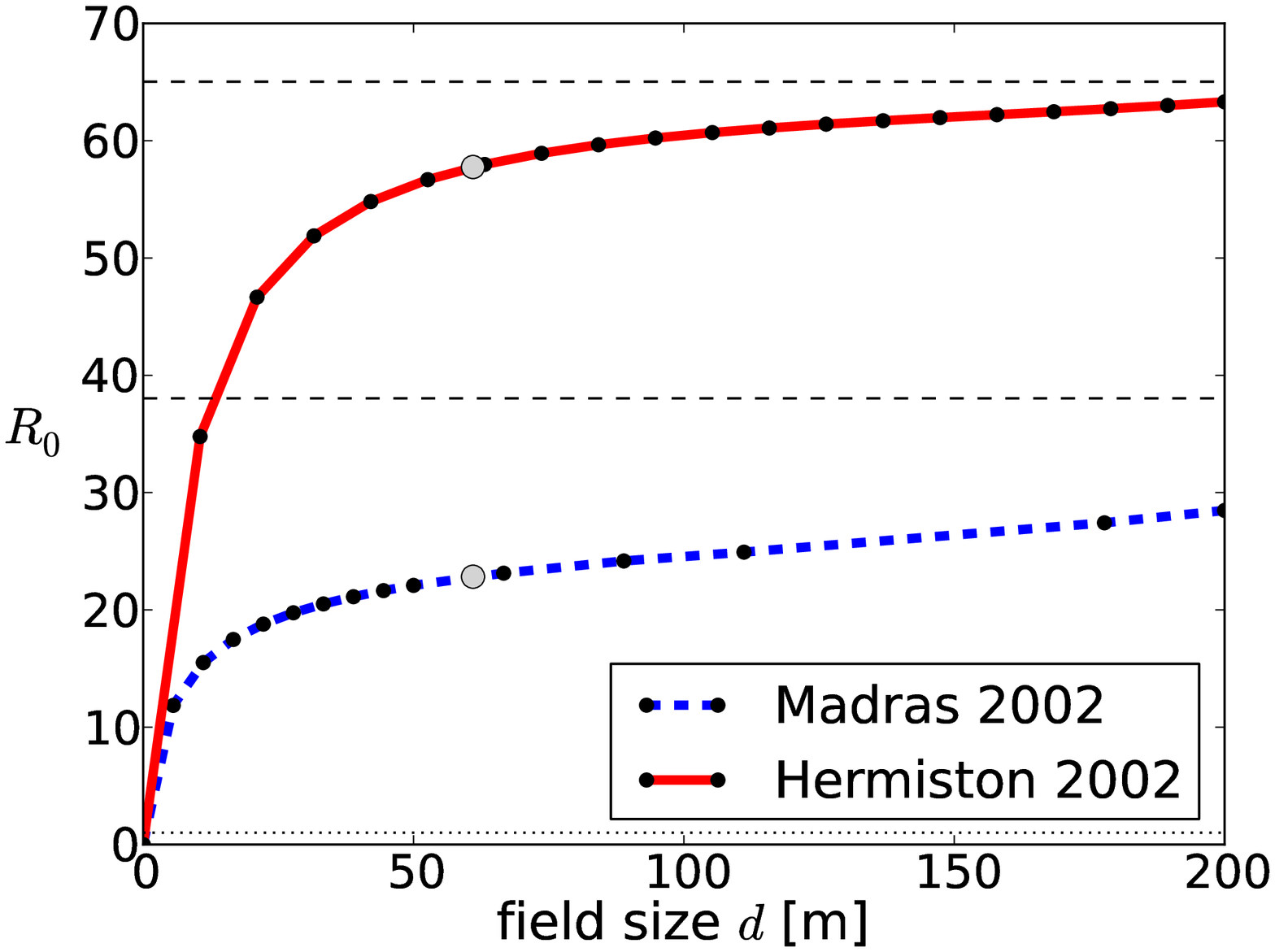}} \caption{
    Basic reproductive number $R_0$ as a function of the field size
    $d$ of a square field calculated [by solving numerically the
    eigenvalue problen \eq{eq:fredh2eigval}] using the modified
    power-law dispersal kernel [\eq{eq:mpowlaw-kern2}] fitted in
    \fig{fig:disgrad-data-fit-herm02} to disease gradient
    datasets (i) Hermiston 2002 downwind (solid curve), and (ii)
    Madras 2002 downwind (dashed curve) obtained in
    \citep{samu05,cowa+05}. Horizontal dashed lines show the
    asymptotic values of the basic reproductive number at large field
    sizes, $R_{0\infty}$, for Hermiston 2002 (upper line) and Madras
    2002 (lower line) datasets. Grey circles indicate the $R_0$-values
    for the field size ($61\,\mrm{m} \times 61\,\mrm{m}$) used in the
    experiments \citep{samu09}.}
\label{fig:r0-vs-a-empir-dprogr}
\end{figure}%
%

\newpage

\begin{figure}
  \centerline{\includegraphics[width=0.5\textwidth]{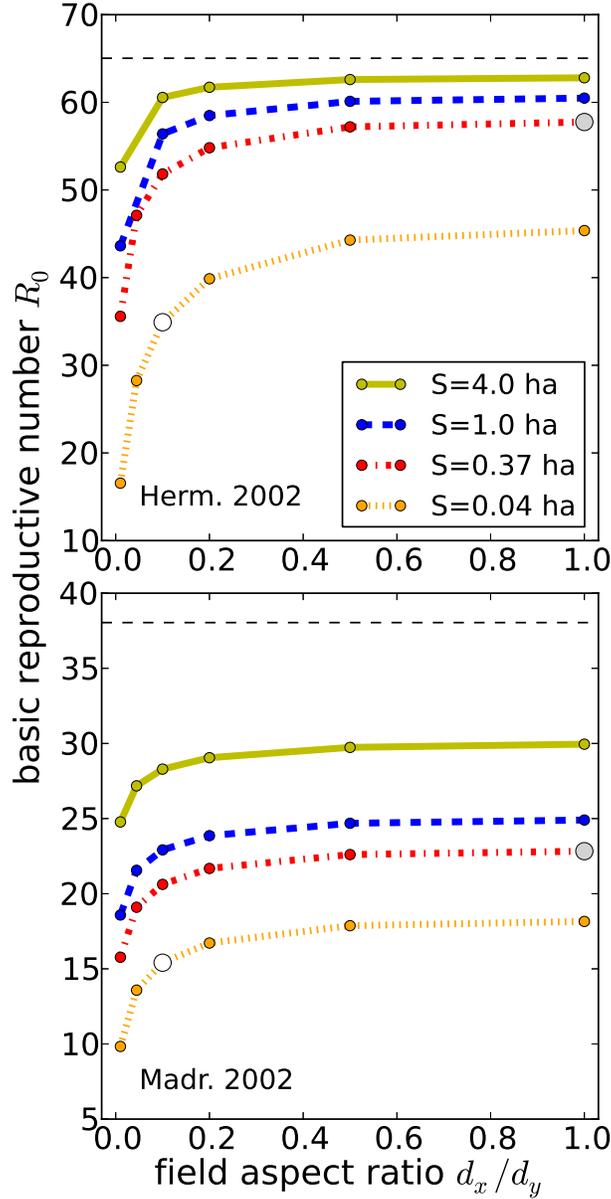}} \caption{Basic
    reproductive number $R_0$ as a function of the field aspect ratio
    $d_x/d_y$ (the field area $S = d_x d_y$ was kept the same). The
    calculation was performed numerically using the power-law
    dispersal kernels fitted to disease gradient data
    (\fig{fig:disgrad-data-fit-herm02}) from Hermiston 2002 (upper
    panel) and Madras 2002 (lower panel) datasets obtained in
    \citep{samu05,cowa+05}. Different curves show the $R_0$ for
    different field areas: $S = 4$\,ha (yellow solid), $S =1$\,ha
    (blue dashed), $S = 0.37$\,ha (red dash-dotted), $S = 0.04$\,ha
    (orange dotted). Larger circles mark the parameters at which the
    field experiments \citep{samu09} were performed (grey circles for
    $6.1\,\mrm{m} \times 61\,\mrm{m} $ and white circles for $61\,\mrm{m} \times
    61\,\mrm{m}$).}
\label{fig:r0-vs-arat-herm02dw}
\end{figure}%
%

%
\end{flushleft}

\appendix
\renewcommand\thefigure{\thesection.\arabic{figure}}  
\numberwithin{equation}{section}

\section{Supporting Information}

\subsection{Linear stability analysis of the disease-free equilibrium}
\label{apsec:linstab}

\setcounter{figure}{0}    

We linearize the model
Eqs.\,(\ref{eq:host-par-ide-2d-1})-(\ref{eq:host-par-ide-2d-2}) in the
vicinity of the disease-free fixed point $H(x,y,t)=K$, $I(x,y,t)=0$
and obtain the following equations for the small deviations from this
fixed point $\xi(x,y,t)$ and $I(x,y,t)$:
\begin{align}
  &\frac{\pd \xi(x, y, t)}{\pd t} = -r_H \xi(x, y, t) -
  \beta K \int \kappa (x,y,u,v) I(u, v, t) du\, dv, \label{eq:dev-ide-2d-1} \\
  &\frac{\pd I(x, y, t)}{\pd t} = \beta K \int \kappa (x,y,u,v) I(u, v, t) du\, dv  - \mu I(x, y, t).  \label{eq:dev-ide-2d-2}
\end{align}
The disease-free fixed point becomes unstable if the small deviation
$I(x,y,t)$ grows over time. To check this, we substitute $I(x,y,t)=
w(x,y) e^{\lambda t}$ in \eq{eq:dev-ide-2d-2}. Then, the stability of
the disease-free fixed point is determined by solving eigenvalue
problem
\begin{equation}\label{eqap:fredh2eigval}
\frac{\beta K}{\mu} \int_0^{d_x} du\, \int_0^{d_y} dv \, \kappa (r) w(u,v) = \sigma w(x,y),
\end{equation}
where $\sigma = 1 + \lambda/\mu$. The eigenvalue problem here consists
in finding the values of $\lambda_j$ and functions $w(x,y)$ satisfying
the relationship (\ref{eqap:fredh2eigval}). The disease-free fixed
point is unstable if at least one of $\lambda_j$ has a positive real
part. \eq{eq:fredh2eigval} is the homogeneous Fredholm equation of the
second kind and can be solved numerically using the Nystrom method
\citep{prte+92}. The dominant eigenvalue $\sigma_d$ determines the
basic reproductive number, i.\,e. $R_0 = \sigma_d$.
Note that the eigenvalue problem \eq{eqap:fredh2eigval} also determines
the stability properties of the corresponding integro-difference
system of equations in discrete time. 

\subsection{Approximation for the basic reproductive number}
\label{apsec:r0-est}

Approximate expression for the basic reproductive number for the model
Eqs.\,(\ref{eq:host-par-ide-2d-1})-(\ref{eq:host-par-ide-2d-2}) can be
found by applying its intuitive definition ``the average number of
secondary cases of infection generated by one primary case in a
susceptible host population'' \citep{anma86} with the averaging
performed over the spatial coordinates. This leads to the
expression:
\begin{equation}\label{eq:naive-r0}
R_\mrm{0c}(x_0,y_0) = \frac{\beta K}{\mu} \int_0^{d_x} dx \int_0^{d_y} dy \, \kappa (x,y,x_0,y_0).
\end{equation} 
Here, the basic reproductive number depends on the position $x_0$,
$y_0$ of the initial inoculum. The basic reproductive number in \eq{eq:naive-r0} does
not yield the invasion threshold at $R_\mrm{0c}(x_0,y_0)=1$ \citep{dihe+90}. However it may
serve as a useful approximate expression, since the calculation
according to \eq{eq:naive-r0} is often much simpler than the solution
of the eigenvalue problem \eq{eqap:fredh2eigval}. 
In order to determine how good this approximation is, we obtain an
explicit expression for $R_\mrm{0c}(x_0,y_0)$
\begin{equation}\label{eq:r0-xy}
R_\mrm{0c}(x_0,y_0) = \frac{\beta K}{4 \mu} \left[ \erf \left( \frac{d_x-x_0}{\sqrt{2}d} \right) + \erf \left( \frac{x_0}{\sqrt{2}d} \right) \right] 
\left[ \erf \left( \frac{d_y-y_0}{\sqrt{2} d} \right) + \erf \left( \frac{y_0}{\sqrt{2}d}\right) \right],
\end{equation}
where we substituted $\kappa(r)$ in \eq{eq:naive-r0} with the Gaussian
dispersal kernel 
\begin{equation}\label{eqap:gauss-pdf}
\kappa_\mrm{G}(r) = \kappa_\mrm{0G} \exp [ -(r/ a)^2].
\end{equation} 

The approximate basic reproductive number $R_\mrm{0c}(x_0, y_0)$ in
\eq{eq:naive-r0} depends on the position of the initial inoculum x0,
y0. In order to obtain a single quantity for a particular spatial
configuration of the host population, we average $R_\mrm{0c}(x_0,
y_0)$ over all possible values of $x_0$, $y_0$ within the field:
\begin{equation}\label{eq:r0-xy-av}
\langle R_{0c}(x_0,y_0) \rangle _{x,y} =  \int_0^{d_x} dx \int_0^{d_y} dy \, R'_0(x,y).
\end{equation}
In the case of the Gaussian dispersal kernel the \eq{eq:r0-xy-av}
yields:
\begin{align}\label{eq:r0-xy-av-res}
\langle R_{0c}(x_0,y_0) \rangle _{x_0,y_0} =  \frac{d^2}{d_x d_y} \frac{\beta K}{\mu}
& \left( \sqrt{\frac{2}{\pi}} (\exp[ -d_x^2/(2 a^2)] - 1) + \frac{d_x}{a}
  \erf \left[ \frac{d_x}{ \sqrt{2} a} \right] \right) \times \\
& \left( \sqrt{\frac{2}{\pi}} (\exp[ -d_y^2/(2 a^2)] - 1) +  \frac{d_y}{a} \erf \left[ \frac{d_y}{ \sqrt{2} a} \right] \right).
\end{align}

\begin{figure}
  \centerline{\includegraphics[width=0.8\textwidth]{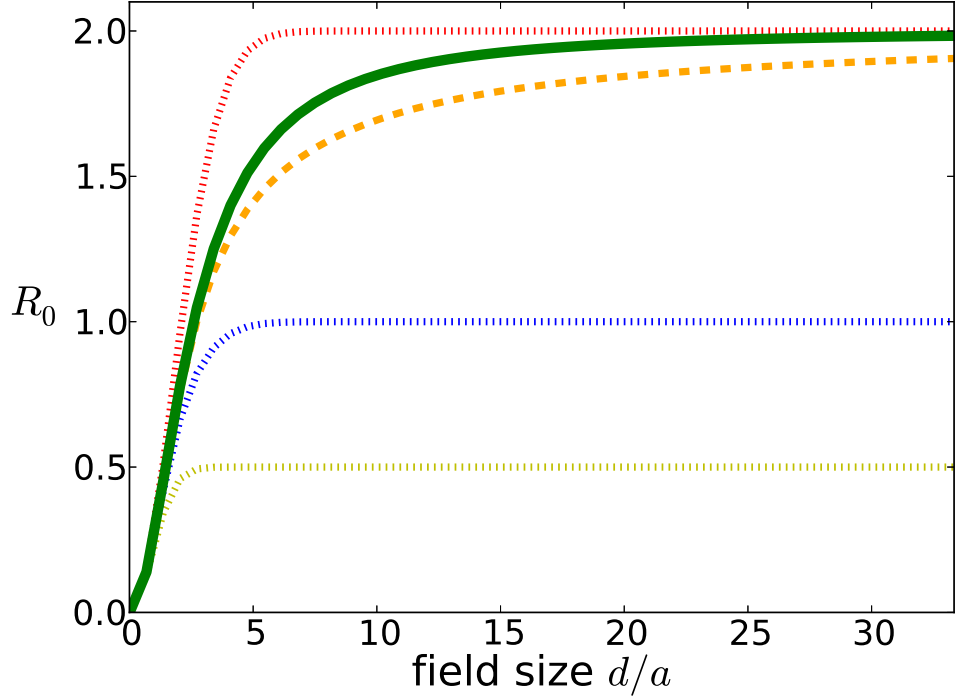}} \caption{
    Basic reproductive number $R_0$ as a function of the field size
    $d$ of the square two-dimensional field measured in units of
    the dispersal radius for the Gaussian dispersal kernel
    [\eq{eqap:gauss-pdf}].  Solid curve shows the $R_0$ computed by
    solving the eigenvalue problem in \eq{eqap:fredh2eigval}. Dotted
    curves represent the approximate $R_{0c}(x_0, y_0)$, according to
    \eq{eq:r0-xy} with the initial inoculum located at the field
    center ($x_0=y_0=d/2$, upper curve), at the field border
    ($x_0=d/2$, $y_0=0$, middle curve) and in the corner of the field
    ($x_0=0$, $y_0=0$, lower curve). The dashed curve shows the
    average $\langle R_{0c}(x_0,y_0) \rangle _{x_0,y_0}$ over the
    field, according to \eq{eq:r0-xy-av-res}. Model parameters:
    $\beta=4$, $K=1$, $\mu=2$.}
\label{fig:r0-vs-a-naive-vs-proper}
\end{figure}

In Figure \ref{fig:r0-vs-a-naive-vs-proper}, the approximate basic
reproductive numbers $R_\mrm{0c}(x_0,y_0)$ calculated using
\eq{eq:r0-xy} (dotted curves), the spatially averaged $\langle
R_{0c}(x_0,y_0) \rangle _{x_0,y_0} $ [\eq{eq:r0-xy-av-res}, dashed
curve] and the exact basic reproductive number obtained by solving
\eq{eqap:fredh2eigval} (solid curve) are plotted versus the field size
$d$. 
The approximate $R_\mrm{0c}(x_0,y_0)$ is highest when the initial
inoculum is introduced to the center of the field (upper dotted curve
in \fig{fig:r0-vs-a-naive-vs-proper}) and is lower at the field border
and in its corner (middle and lower dotted curves in
\fig{fig:r0-vs-a-naive-vs-proper}).
The spatial averaged $\langle R_\mrm{0c}(x_0,y_0) \rangle _{x_0,y_0}$
is reasonably close to the actual $R_0$ (cf. dashed and solid curves
in \fig{fig:r0-vs-a-naive-vs-proper}), but it underestimates the
actual $R_0$, because it neglects the contribution of the subsequent
generations of infection. 
At $d \gg a$ the $R_0$ tends asymptotically to the maximal value of
$R_\mrm{0c}(x_0,y_0)$, achieved at the field center $x=d/2$,
$y=d/2$. The values of $R_\mrm{0c}(x_0,y_0)$ at the border and in the
corner of the field also reach constant but considerably smaller
values at $d \ll a$. This can be explained by the fact that when the
size of the field increases, the surface-to-volume ratio of the
square field decreases, meaning that the contribution of the hosts
close to the field border to $R_0$ steadily decreases.
 

%

All the curves in \fig{fig:r0-vs-a-naive-vs-proper} behave in the same
way at small field sizes (i.\,e. when $d \ll a$): they increase
quadratically with the field size $d$, according to
\begin{equation}\label{eq:r0-asympt-sm-a}
R_\mrm{0asympt} = \frac{\beta K}{2 \pi a^2 \mu} d^2.
\end{equation}
Thus, the approximate expression for the basic reproductive number
\eq{eq:naive-r0} holds well in the two limiting cases: at small field
sizes (i.\,e. when $d \ll a$) and at large field sizes (i.\,e. when $d
\gg a$).

\subsection{Estimation of the basic reproductive number as a function
  of the field size and shape}
\label{apsec:estim-r0-vs-a-arat}

The basic reproductive number, $R_0$ can be determined as the dominant eigenvalue of
the Fredholm equation \eq{eqap:fredh2eigval}
We compute it as a function of the dimensions $d_x$ and $d_y$ of a
rectangular field, which characterize its size and shape. To do this,
we obtain numerical estimates for the dispersal kernel $\kappa(r)$
(\sec{apsec:fit-dg} and \sec{apsec:dispfung-def-norm}) and the
parameter combination $\beta K / \mu$ (\sec{apsec:estim-r0-inf}), which as
we will show corresponds to the limit of $R_0$ at $d_x,\, d_y \to
\infty$.

\subsubsection{Fitting disease gradients}
\label{apsec:fit-dg}



\begin{figure}
  \centerline{\includegraphics[width=0.8\textwidth]{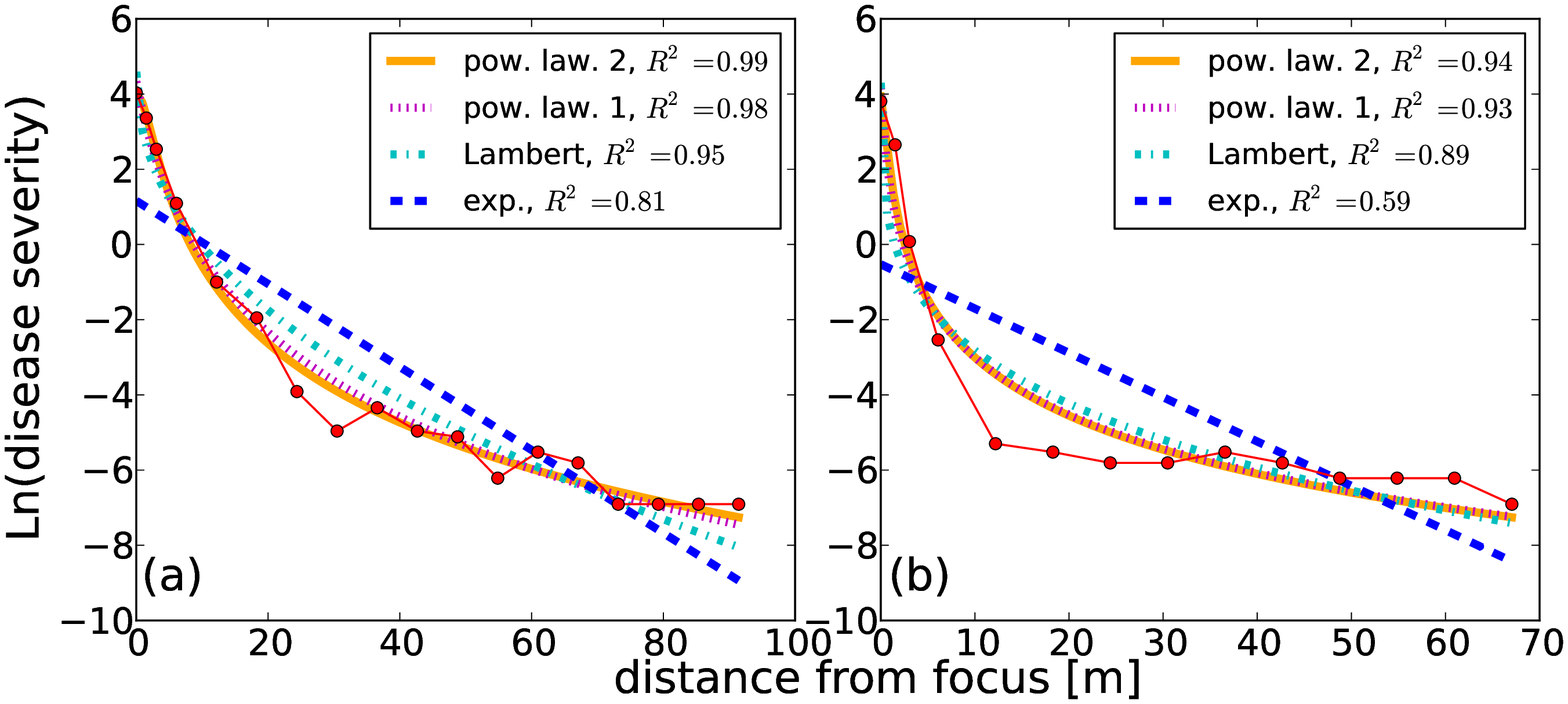}}\caption{Disease
    gradient data (circles) from Hermiston 2002 downwind [left panel
    (a)] and Madras 2002 downwind [right panel (b)] experiments
    conducted by \citet{samu05,cowa+05}. Natural logarithm of
    disease severity is shown versus the distance from focus. The data
    was fitted by four functions: exponential [\eq{eqap:lamb-kern}
    with $n=1$], Lambert [\eq{eqap:lamb-kern}], modified power law 1
    [\eq{eqap:mpowlaw-kern1}] and modified power law 2
    [\eq{eqap:mpowlaw-kern2}].}
\label{figap:disgrad-data-fit-herm02-madr02}
\end{figure}

Disease gradients were measured in terms of both average number of
lesions per leaf and disease severity in a large-scale experiment over
three consecutive seasons \citep{samu05,cowa+05}. The datasets
corresponding ot the average numbers of lesions per leaf in primary
disease gradients were fitted using several different model functions
\citep{samu05}. Here, we also fitted the disease severity measurements
corresponding to primary disease gradients
(\fig{figap:disgrad-data-fit-herm02-madr02}) for the two largest
datasets (Hermiston 2002 and Madras 2002) of the experiments
\citep{samu05,cowa+05}.

The following model functions are often used to fit the disease
gradient data.
Lambert kernel \citep{lavi+80} 
\begin{equation}\label{eqap:lamb-kern}
y_\mrm{L}(r) = y_{0} \exp [ -(r/ a)^n],
\end{equation} 
%
%
%
%
which includes the special cases of the exponential (or Laplacian)
kernel at $n=1$ and the Gaussian kernel at $n=2$.
Power-law kernel \citep{gr68} 
\begin{equation}\label{eqap:powlaw-kern}
y_\mrm{PL}(r) = y_0 r^{-b}
\end{equation} 
is used to describe disease gradients of pathogens with long-range
dispersal. However, the function approaches infinity at the focus
$r=0$, which is unrealistic. For this reason a modified power-law
kernel 
was introduced \citep{mule85}
\begin{equation}\label{eqap:mpowlaw-kern1}
y_\mrm{PL1}(r) = y_\mrm{0} (r_0 + r)^{-b}.
\end{equation} 
It exhibits the same behavior as the power-law kernel in
\eq{eqap:powlaw-kern} at large $r$, but the divergence is ``softened'' such
that the function has a finite value at $r=0$. In this study, we
used a different form of the modified power-law kernel
\begin{equation}\label{eqap:mpowlaw-kern2}
y_\mrm{PL2}(r) = y_0 \left( r_0^2 + r^2 \right)^{-b/2}
\end{equation} 
that is very similar to \eq{eqap:mpowlaw-kern1}, but is more suitable for
extensive numerical computations required for the solution of the
eigenvalue problem in \eq{eq:fredh2eigval}.

Figure\,\ref{figap:disgrad-data-fit-herm02-madr02} shows the primary
disease gradients in terms of the disease severity for the two largest
datasets obtained in \citep{cowa+05,samu05}: Hermiston 2002 (left
panel) and Madras 2002 (right panel). Both of the datasets were fitted
using the exponential kernel [\eq{eqap:lamb-kern} with $n=1$], Lambert
kernel [\eq{eqap:lamb-kern}], modified power law 1
[\eq{eqap:mpowlaw-kern1}] and modified power law 2
[\eq{eqap:mpowlaw-kern2}]. The two modified power laws provided best
fits with the modified power law 2 being slightly better. It is
our kernel of choice, since it also allows for faster numerical
solutions of the eigenvalue problem in \eq{eq:fredh2eigval}.

 

The fit of the modified power-law function in \eq{eqap:mpowlaw-kern2}
to the disease gradient data shown in
\fig{figap:disgrad-data-fit-herm02-madr02} yielded the following
estimates for the parameter values:
\be\label{eq:params-herm02dw}
\mrm{Hermiston\:2002\:downwind}\:r_0 = 2.2255\:\mrm{\,m,}\:b = 3.0365,\:y_0 =
6.4424;
\ee
\be\label{eq:params-madr02dw}
\mrm{Madras\:2002\:downwind}\: r_0 = 0.4486\mrm{\,m,}\:b = 2.2345, y_0 = 0.085127.
\ee

%

\subsubsection{Definition and normalization of the dispersal kernel}
\label{apsec:dispfung-def-norm}

We defined the dispersal kernel $\kappa (x,y,u,v) $ as a probability density
function for an infectious spore to land at a distance $r$ from its
source \citep{nakl+12}. A spore should eventually land somewhere is
reflected in the condition to normalize the dispersal kernel:
\begin{equation}\label{eq:kappa-norm-cond}
\int_0^{2 \pi} d \theta \int_0^\infty d r r \kappa(r, \theta) = 1.
\end{equation}
Here, we transformed the dispersal kernel to polar coordinates using
the relationships $x = r \cos \theta$, $y = r \sin \theta$. In the
case of isotropic dispersal $\kappa(r, \theta) = \kappa(r)$, i.\,e. the
dispersal kernel does not depend on the angle of dispersal
$\theta$. Then the normalization condition reads
\begin{equation}\label{eq:kappa-norm-cond-isotr}
2 \pi \int_0^\infty d r r \kappa(r) = 1.
\end{equation}


Next, we provide the normalization condition for the modified
power-law function $Y_\mrm{PL2}(r)$ [\eq{eqap:mpowlaw-kern2}] and for
the Lambert function [\eq{eqap:lamb-kern}].


The dispersal kernel $\kappa(r)$ is assumed to be proportional to the
disease gradient $y(r)$ (see \sec{sec:case-pstriiformis}). Therefore,
the dispersal kernel should be given by the same function as the
disease gradient
\begin{equation}\label{eqap:mpowlaw2-pdf}
\kappa_\mrm{PL2}(r) = \kappa_\mrm{0PL2} \left( r_0^2 + r^2 \right)^{-b/2},
\end{equation} 
but with the different proportionality constant $\kappa_0$, which is
obtained by substituting the \eq{eqap:mpowlaw2-pdf}  into the normalization condition \eq{eq:kappa-norm-cond-isotr}:
\be\label{eq:kappa0-pl2}
\kappa_\mrm{0PL2}  = (b - 2) r_0^{b-2}/(2 \pi).
\ee
This expression is valid only if the integral in
\eq{eq:kappa-norm-cond-isotr} converges, which is the case at
$b>2$. In both datasets used here (Hermiston 2002 and Madras 2002
downwind) this condition is fulfilled for the values of $b$,
corresponding to the best fit.

Similarly, the Lambert dispersal kernel has the form:
\begin{equation}\label{eqap:lamb-pdf}
\kappa_\mrm{L}(r) = \kappa_\mrm{0L} \exp [ -(r/ a)^n],
\end{equation} 
where 
\be\label{eq:kappa0-lamb}
\kappa_\mrm{0L} =\frac{1}{\pi a^2 \Gamma \left( \frac{2+n}{n} \right)} 
\ee
is determined from the normalization condition \eq{eq:kappa-norm-cond-isotr}.

We use the numerical values for the best-fit parameters
\eq{eq:params-herm02dw} and \eq{eq:params-madr02dw} to obtain
estimates for $\kappa_0$ using \eq{eq:kappa0-pl2}:

\be\label{eq:herm02dw-kappa0-pl2}
\mrm{Hermiston\:2002\:downwind}:\:\kappa_0 = 0.3780,\:
\ee
\be\label{eq:madr02dw-kappa0-pl2}
\mrm{Madras\:2002\:downwind}:\:\kappa_0 = 0.03092.
\ee

Thus, our estimates for the dispersal kernels $\kappa(r)$ 
are given by the \eq{eqap:mpowlaw2-pdf} with the parameter values from
\eq{eq:params-herm02dw} and \eq{eq:herm02dw-kappa0-pl2} for Hermiston
2002 downwind; and from \eq{eq:params-madr02dw} and \eq{eq:madr02dw-kappa0-pl2}.

\subsubsection{Estimation of the $R_0$ in the limit of a large field size}
\label{apsec:estim-r0-inf}

First, we consider the host population to be initially fully susceptible and
have the leaf area index of $K_0$. Then, we introduce a localized unit of
infected hosts (focus) at a position $x_0$, $y_0$
\begin{equation}\label{eq:unit-init-infect1}
H(x,y,t=0)=K, \: I(x, y, t=0) = I_\mrm{tot0} \delta(x-x_0) \delta(y-y_0).
\end{equation}
%
%
We are interested here only in the primary infections occuring due
$I(x, y, t=0)$, because the amount of disease due to the primary
infection (or the primary disease gradient) is often measured in
experiment (for example, \citep{samu05}).  Hence, we derive the amount
of infection produced after a single time step $\Delta t$ from
\eq{eq:host-par-ide-2d-2}:
\begin{align}\label{eq:infects-td}
& \left[ I(x, y, t=\Delta t) - I(x, y, t=0) \right] / \Delta t = \\
& \beta \int_0^{d_x} du \int_0^{d_y} dv \kappa(x,y,u,v) I(u,v,t=0) H(x,y,t=0) - \mu I(x,y,t=0)
\end{align}

By substituting \eq{eq:unit-init-infect1} in \eq{eq:infects-td} we
obtain
\begin{equation}\label{eq:infects-td1}
 \Delta I(x, y, t=\Delta t) = I_\mrm{tot0} \Delta t \, K_{\Delta t} \beta \kappa(x,y,x_0,y_0),
\end{equation}
where 
\be\label{eq:deltai}
\Delta I(x, y, t=\Delta t) = I(x, y, t=\Delta t) - I(x, y, t=0)
\ee
 represents the primary disease gradient from a localized point-like
source. 
Further, we assume dispersal to be isotropic and set the coordinate of
the focus to zero, i.\,e. $x_0=0$. Then, the amount if infected host
in the next time step and the dispersal function depend only on the
distance $r = \sqrt{x^2 + y^2}$ from the focus, i.\,e. $I(x, y, t=\Delta t) = I(r, t=\Delta
t)$,  $\kappa(x,y,x_0,y_0) = \kappa(r)$. We can then re-write the
\eq{eq:infects-td1}:
\begin{equation}\label{eq:infects-td2}
 \Delta I(r, t=\Delta t) = I_\mrm{tot0} \Delta t \, K_{\Delta t} \beta \kappa(r),
\end{equation}


Next, we connect $\Delta I(x, y, t=\Delta t)$ with the whole-plant
disease severity $y(r)$.

The quantity $I(r, t)$ in our model that represents the spatial
density of the infected host tissue. In the case of wheat stripe rust
it is the infected leaf area per unit land area (in analogy with the
``leaf area index'' (LAI), we will call it the ``infected leaf area
index'' (ILAI)). We express the disease severity as a ratio $y(r)
= \mathcal{I}(r)/ \mathcal{K}_{\Delta t}$, where $\mathcal{I}(r)$ is
the total infected leaf area at a location $r$ and
$\mathcal{K}_{\Delta t}$ is the total leaf area at a location. By
dividing both the numerator and the denominator of this expression by
the unit land area $\Delta s$, we obtain $y(r) = \Delta
I(r)/K_{\Delta t}$, where $\Delta I(r)$ is given by \eq{eq:deltai},
and $K_{\Delta t}$ is the total leaf area index. Therefore,
\begin{equation}\label{eq:deltai-propto-sev}
\Delta  I(r, t) =\Delta t) = K_{\Delta t} y(r).
\end{equation}
On the other hand, from \eq{eq:infects-td2}
\begin{equation}\label{eq:deltai-propto-kappa}
\Delta I(r, t=\Delta t) = \beta K_{\Delta t} \Delta t   I_\mrm{tot0} \kappa(r).
\end{equation}
%
By equating \eq{eq:deltai-propto-sev} and \eq{eq:deltai-propto-kappa}
we obtain the relationship
\begin{equation}\label{eq:disp-func-prop-disgrad}
\frac{\beta}{\mu} = \frac{1} {I_\mrm{tot0} } \frac{y(r)} {\kappa(r)}.
\end{equation}
Here we assumed $\Delta t = 1/\mu$, which implies that the consecutive
pathogen generations do not overlap (see the discussion in \sec{apsec:linstab}).
We multiply both sides of the \eq{eq:disp-func-prop-disgrad} by the
leaf area index $K_{\Delta t}$ at time $t = \Delta t$ and obtain the
expression for $R_{0\infty} = \beta K_{\Delta t} /\mu$

\begin{equation}\label{eq:r0infty-vs-params}
R_{0\infty} = \frac{K_{\Delta t}} {I_\mrm{tot0}} \frac{Y_0}
{\kappa_0}.
\end{equation}
Here we used the fact that $\kappa(r)$ is proportional to $Y(r)$ and,
therefore, their ratio equals to the ratio $Y_0/\kappa_0$.

Now, we determine the intensity of the initial inoculum $I_\mrm{tot0}$
[\eq{eq:unit-init-infect1}] from experimental parameters.
The $\delta$-functions in \eq{eq:unit-init-infect1} represent an
infinitely narrow peak of a unit height. This is an idealized
mathematical entity that can, however, be quite useful. It describes
the actual situation well if the spatial scale of interest is much
larger than the size of the focus. This was the case in the studies
\citep{samu05,cowa+05}, where the focus (the area inoculated initially)
was a square with the side $\Delta x_f = 1.52$\,m, while the spatial
scale over which the epidemic developed in the next generation was
50-80\,m for the two largest datasets (Hermiston 2002 and Madras 2002
downwind). 
\begin{equation}\label{eq:cnct-itot0-sev0}
\int_0^{\Delta x_f} d x \int_0^{\Delta x_f} d y I_\mrm{tot0}
\delta(x-x_0) \delta(y-y_0) = I_\mrm{tot0} = \int_0^{\Delta x_f} d x
\int_0^{\Delta x_f} d y I_0 = y_0 K_0 \Delta x_f^2.
\end{equation}
Here, $y_0$ is the disease severity at the focus caused by
artificially inoculated spores (first generation) and $K_0$ is the
leaf area index at the time of inoculation (``zeroth''
generation). The \eq{eq:cnct-itot0-sev0} says what the intensity of
the initial inoculum should be if it was concentrated in a very small
area such that the total amount of disease is the same as in the
experiment. 
\begin{equation}\label{eq:cnct-itot0-sev0-shrt}
I_\mrm{tot0} = y_0 K_0 \Delta x_f^2.
\end{equation}
After substituting \eq{eq:cnct-itot0-sev0-shrt} into
\eq{eq:r0infty-vs-params} we obtain:
\begin{equation}\label{eq:r0infty-vs-params1}
R_{0\infty} = \frac{K_{\Delta t}} {K_0} \frac{1} {y_0 \Delta x_f^2} \frac{Y_0}
{\kappa_0}.
\end{equation}
The expression in \eq{eq:r0infty-vs-params1} now consists of the
parameters that are known from a typical disease gradient experiment.

We use the estimates we obtained above for the parameters $Y_0$
[\eq{eq:params-herm02dw} and \eq{eq:params-madr02dw}] and $\kappa_0$
[\eq{eq:herm02dw-kappa0-pl2} and \eq{eq:madr02dw-kappa0-pl2}], also
use the area of the focus $\Delta x_\mrm{f}^2 =1.52\,\mrm{m} \times
1.52\,\mrm{m}=2.31\,$m$^2$ for both datasets and the values for the
initial disease severity $y_0=0.227$ (Hermiston 2002) and $y_0=0.062$
(Madras 2002) \citep{cowa+05}. We also assume that the leaf area index
at the time of inoculation $K_0$ was two times smaller than its value
at the time of disease gradient measurement, when the plants almost
reached their maximum size, i.\,e. $K_{\Delta t} = 2 K_0$.  By
substituting these values into \eq{eq:r0infty-vs-params1} we obtain
the estimates for $R_{0\infty}$:
\be\label{eq:r0-inf-herm02dw}
\mrm{Hermiston\:2002\:downwind}\:R_{0\infty} = 65.0;
\ee
\be\label{eq:r0-inf-madr02dw}
\mrm{Madras\:2002\:downwind}\:R_{0\infty} = 38.0.
\ee

 
Having obtained the numerical values for the parameter $R_{0\infty} =
\beta K_{\Delta t} / \mu$ and the function $\kappa(r)$, we solved the
eigenvalue problem in \eq{eqap:fredh2eigval} numerically for different
values of $d_x$ and $d_y$ and determined the basic reproductive number
$R_0$ as a function of the field size and shape. The results of this
computation are shown in \fig{fig:r0-vs-a-empir-dprogr} and
\fig{fig:r0-vs-arat-herm02dw}.


\subsection{Susceptible-infected model with spatial spore dispersal}
\label{apsec:si-spore-disp}

In this section we consider the model that takes into account spore
dynamics explicitly. Our goal here is to describe the approximation
that was used to obtain the simplified model
Eqs.\,(\ref{eq:host-par-ide-2d-1})-(\ref{eq:host-par-ide-2d-2}) that
do not explicitly include spore dynamics. For the sake of brevity we
consider the model in one-dimensional space, but it is straightforward
to extend the consideration to two dimensions.
The model of host-pathogen population dynamics reads
\begin{align}
\frac{\pd H(x,t)}{\pd t} &= r_H ( K - H(x,t)) - \beta' \int_0^d \kappa(|s-x|)
U(s,t) ds \, H(x,t), \label{eq:1host1fung1spore-1d-1}\\ 
\frac{\pd I(x,t)}{\pd t} &= \beta' \int_0^d \kappa(|s-x|) U(s,t) ds \, H(x,t) -\mu I(x,t), \label{eq:1host1fung1spore-1d-2}\\ 
\frac{\pd U(x,t)}{\pd t} & = \gamma I(x,t) - \mu' U(x,t), \label{eq:1host1fung1spore-1d-3}
\end{align}
where $H(x,t)$, $I(x,t)$ represent the areas covered by susceptible
and infected host tissue, correspondingly, per unit area of the field;
and $U(x,t)$ represents the number of spores per unit area of the
field.
Susceptible hosts $H(x,t)$ grow with the rate $r_H$. Their growth
is limited by the ``carrying capacity'' $K$, implying limited space or
nutrients. 
Furthermore, susceptible hosts $H(x,t)$ may be infected by the
pathogen and transformed into infected hosts in the compartment
$I(x,t)$ with the transmission rate $\beta'$. The corresponding terms
in
Eqs.\,(\ref{eq:1host1fung1spore-1d-1})-(\ref{eq:1host1fung1spore-1d-2})
are proportional to the amount of the available susceptible tissue
$H(x,t)$ and to the amount of the infectious spores $U(x,t)$ at the
location $x$.  Infectious spores are produced at the rate $\gamma$ and
lost at the rate $\mu'$.

 Here, $\kappa(|s-x|)$ is the dispersal kernel that
characterizes the probability of an infectious spore, produced at the
location $s$ to land at the location $x$. The integration is performed
over all possible sources of spores within the field, i.\,e. over the
whole extension of the field from 0 to $d$, where $d$ is the size of
the field. We assume that the dispersal kernel depends only on the
distance $|s-x|$. The fact that the spore should land somewhere allows
to normalize this function such that the integral of it over the whole
space is unity:
\begin{equation}\label{eq:kappa-norm}
\int_0^{\infty} \kappa(r) J(r) dr = 1,
\end{equation}
where $J(r)=1$ for the one-dimensional case considered here, and $J(r)=r$ for the
two-dimensional case (in this case additional integration over the
polar angle is required).

We assume that the characteristic time scale of spore dispersal is
much shorter than the characteristic time scales associated with other
stages of the pathogen life cycle. Then, the equation for spores is
assumed to quickly assume the equilibrium state, with the left-hand
side equal to zero and $U(x,t) = (\gamma / \mu') I(x,t)$. This means
that the density of spores is proportional the density of the
infectious host tissue. By substituting this expression into
Eqs.\,(\ref{eq:1host1fung1spore-1d-1})-(\ref{eq:1host1fung1spore-1d-3}),
we reduce the model to just two
Eqs.\,(\ref{eq:host-par-ide-2d-1})-(\ref{eq:host-par-ide-2d-2}), where
the transmission rate is a compound parameter: $\beta = \gamma \beta'
/ \mu'$.

\subsection{The relationship between the basic reproductive number and
the epidemic velocity}
\label{apsec:relat-r0-epvel}


For the susceptible-infected epidemiological model where the
transmission of disease through space is described using the diffusion
term (proportional to the Laplacian of $I(x,y,t)$), the wave speed of
the epidemic, $c$, is proportional to $\sqrt{R_0 - 1}$
\citep{kero08epvel}. This relationship holds in the case of very local
dispersal: the diffusion term can be obtained from a more general
formulation in terms of a system of integro-differential equations by
performing the Taylor series expansions under the assumption that the
dispersal is sufficiently local. In addition, this requires that the
average dispersal distance is finite, and hence the dispresal kernel
must decay faster than $r^{-3}$.

In our case the dispersal is nonlocal and is governed by empirically
determined dispersal kernels that exhibit power-law behavior. In this
case, it is not straightforward to determine the analytical relationship between
the basic reproductive number and the epidemic velocity. A numerical
investigation can be performed by solving the system of
Eqs.\,(\ref{eq:host-par-ide-2d-1})-(\ref{eq:host-par-ide-2d-2})
numerically with the parameters corresponding to different values of
$R_0$ and determining the epidemic velocity. 

However, we can still use the relationship $c \propto \sqrt{R_0 - 1}$
as a rough lower estimate for the epidemic velocity in this
case. Then, the ratio between the epidemic velocities $c_1$ and $c_2$
in plots with different sizes and geometries reads:
\be\label{eq:epvel-ratio}
\frac{c_1}{c_2} = \sqrt{\frac{R_{01} - 1} {R_{02} - 1}},
\ee
where $R_{01}$ and $R_{02}$ are the basic reproductive numbers in
these two different plots. We obtained the following estimates for the
basic reproductive numbers that correspond to the two plot sizes and
geomteries (plot 1: $61\,\mrm{m}\times 61\,\mrm{m}$; plot 2:
$6.1\,\mrm{m}\times 61\,\mrm{m}$) used in the experiments
\citep{samu09} (these are marked as white and gray circles in
\fig{fig:r0-vs-arat-herm02dw})
\be\label{eq:r0vals-herm02dw}
\mrm{Hermiston\:2002:}\:R_{01} = 57.75,\:R_{02} = 34.91;
\ee
\be\label{eq:r0vals-madr02dw}
\mrm{Madras\:2002:}\:R_{01} = 22.83,\:R_{02} = 15.41;
\ee
Substituting these values in \eq{eq:epvel-ratio} leads to the
following approximate ratios of the epidemic velocities:
\be\label{eq:epvel-rat-herm02dw}
\mrm{Hermiston\:2002:}\:\frac{c_1}{c_2} = 1.286;
\ee
\be\label{eq:epvel-rat-madr02dw}
\mrm{Madras\:2002:}\:\frac{c_1}{c_2} = 1.217.
\ee
Thus, we predict a moderate difference in epidemic velocities in these
two plots, while the empirical study \citep{samu09} reported no
detectable difference. We suggest two possible explantaions for this
discrepancy. First, our model assumed isotropic dispersal and
neglected the influence of the prevailing wind direction, while in the
experimental setting of \citep{samu09}, there was a strong anisotropy
in dispersal due to wind. Strongly directional wind may be capable of
masking the effect of plot size and geometry on $R_0$ and epidemic
velocity. This is because the smaller or narrower plots decrease
pathogen fitness due to the edge effect, i.\,e. due to the pathogen
spores that were lost outside the plot. In the presence of a strong
wind, in an elongated plot, the spores that would have been lost
outside the plot may well remain inside and contribute to the
development of the epidemic. This effect is expected to be strongest
when the prevailing wind direction coincides with the longer axis of
the plot, as was the case in the experimental setting
\citep{samu09}. On the contrary, we expect the effect of the plot size
and geomtry to be magnified by wind, when the wind direction is
perpendicular to the longer axis of the plot.
A second possible factor that may contribute to the discrepancy is the
experimental resolution: it may be challenging to be able to detect
differences in epidemic velocities of 20-30\,\% that we predict in
(\ref{eq:epvel-rat-herm02dw}), (\ref{eq:epvel-rat-madr02dw}).





\end{document}